\documentclass[a4paper,fleqn,usenatbib,useAMS]{mnras}

\usepackage{graphicx}	
\usepackage{amsmath}	
\usepackage{amssymb}	
\usepackage{multicol}        
\usepackage{bm}		
\usepackage{pdflscape}	
\usepackage{subfigure}
\usepackage[usenames]{color}  

\usepackage[T1]{fontenc}
\usepackage{ae,aecompl}

\usepackage{newtxtext,newtxmath}

\title[Spirals in dwarf galaxies]{\textit{On the formation of spiral arms in dwarf galaxies}}

\author[A. V. Zasov]{Anatoly V. Zasov$^{1,2}$\thanks{Contact e-mail: \href{mailto:zasov@sai.msu.ru}{zasov@sai.msu.ru}}, Alexander V. Khoperskov$^{3}$\thanks{Contact e-mail: \href{mailto:khoperskov@volsu.ru
}{khoperskov@volsu.ru}}, Natalia A. Zaitseva$^{2,4}$ \newauthor and 
Sergey S. Khrapov$^{3}$
\\
$^{1}$Sternberg Astronomical Institute, Moscow M.V. Lomonosov State University, Universitetskij pr., 13, Moscow 119234, Russia
\\
$^{2}$Faculty of Physics, Moscow M.V. Lomonosov State University, Leninskie gory 1, Moscow 119991, Russia
\\
$^{3}$Volgograd State University, Universitetsky pr., 100, Volgograd 400062, Russia
\\
$^4$Special Astrophysical Observatory of RAS, Nizhnij Arkhyz, Karachai-Circassia 369167, Russia}

\pubyear{Astronomy Reports, 2021, vol.??, p.??-?? (Accepted for publishing in ARep)}

\begin{document}
\label{firstpage}
\maketitle

\begin{abstract}
Spiral structure (both flocculent and Grand Design types) is very rarely observed in dwarf galaxies because the formation of spiral arms requires special conditions.  In this work we analyze the sample of about 40 dS-galaxies found by scanning by eye the images of late-type galaxies with $m_B<15^m$ and $M_B>-18^m$ and photometric diameter $D_{25}<12$~kpc.  We found that apart from the lower average  gas (HI) fraction the other properties of dS-galaxies including the presence of a bar and the isolation index do not differ much from those for dwarf Irr or Sm-types of similar luminosity and rotation velocity (or specific angular momentum).There are practically no dS-galaxies  with rotation velocity below 50\,--\,60~km\,sec$^{-1}$.

To check the conditions of formation of spiral structure in dwarf galaxies we carried out a series of N-body/hydrodynamic simulations of low-mass stellar-gaseous discy galaxies by varying the model kinematic parameters of discs, their initial thickness, relative masses and scale lengths of stellar and gaseous disc components, and stellar-to-dark halo masses.
We came to conclusion that the gravitational mechanism of spiral structure  formation is effective only for thin stellar discs, which are  non-typical for dwarf galaxies, and for not too slowly rotating galaxies. Therefore, only a small fraction of dwarf galaxies with stellar/gaseous discs have spiral or ring structures. The thicker stellar disc, the more gas is required for the spiral structure to form. The reduced   gas content in many dS-galaxies compared to non-spiral ones may be a result of more efficient star formation due to a higher volume gas density thank to the thinner stellar/gaseous discs.

\end{abstract}




\section{Introduction}
\label{sec:intro}
Usually a spiral structure is not observed in low-luminous galaxies. Most of dwarf galaxies possessing stellar discs belong to Irr galaxies in the  Hubble classification scheme (Im or Sm type in de~Vaucouleurs' classification) with the irregular inner structure (note that the difference between Sm and classical Irr galaxies is rather arbitrary and poorly defined). A spiral structure rarely presents in dwarf galaxies  (See f.e. \citet{Edmunds&Roy93}, \citet{Hidalgo04}).  
However, a small fraction of low luminous galaxies still possess a well distinguishable spiral pattern either of flocculent or (rare) a Grand Design type, and may be ascribed to dS-type. 

Indeed, the existence of long-lived spiral density waves requires certain dynamic conditions, and, first of all, it is necessary for a galaxy  to have a   dynamically cold  stellar or gaseous disc, which evidently rarely occurs in dwarfs.  It is of interest to study the peculiar properties of these galaxies and search for the conditions which may lead to formation of a spiral structure.

A search of dwarf spirals is complicated by their low luminosity and small angular sizes of their discs, which limits the sample of galaxies with a clearly traceable spiral structure to relatively close objects.  In addition, the internal structure in dwarf galaxies is often very noisy with star-forming regions scattered across the disc. 

The first representative statistical study of dwarf spiral galaxies was carried out by \citet{Hidalgo04} by considering a sample of more than one hundred dS-objects. A selection of dS galaxies was performed by Hidalgo-G{\'a}mez mostly by visual selection of Sm galaxies from the UGC catalog and partially of late-type galaxies in \citet{NGC} catalog. A spiral galaxy was considered as a dwarf one if its absolute blue magnitude is higher than $-18^m$. Selected dS-galaxies do not belong to rich clusters except one Virgo galaxy (UGC 7781). By comparing dS and normal high-luminous S-galaxies, the author came to conclusion that dwarf spirals cannot be considered simply as a smaller version of giant ones. They seem to be more similar to dwarf irregular galaxies than to the classical spirals, differing from most dwarf galaxies without spiral structure by higher (on average) luminosity and larger size.  By their spectral characteristics they are indistinguishable from Irr galaxies and also have a low gradient of gas metallicity. In addition, a Hidalgo-G{\'a}mez' sample of dS-galaxies seems to have a deficiency of barred systems compared to giant spiral galaxies (however we do not confirm this result, see below).  

A more detailed comparison of star formation rate of dwarf galaxies possessing spiral arms (dS) and without spirals (Sm) was presented by \citet{MaganaSFR20}. They concluded that Sm galaxies have higher star formation rate values (SFR) than dS, however when the surface density of SFR is considered, both types of galaxies are indistinguishable.

In this work  we compile and investigate a sample of dwarf galaxies with a spiral structure using HyperLeda database (\citet{HyperLeda}). 
A general idea is to compare dS galaxies with the reference sample of non-spiral dwarf discy galaxies of similar luminosity, size, or, when possible, dynamic properties. Then we try to clarify the conditions for the formation and existence of long-lived spiral pattern in low-mass galaxies, using numerical self-consistent 3D models of  galaxies, rotating with the velocities $V^{max}\le 100$\,km\,sec$^{-1}$, typical for dwarf systems.  

The vast majority of dS-galaxies belongs to the late morphological types. Here we ignore a very rare type of stellar spiral arms discovered in dE-galaxies, which are barely distinguishable at a bright background of a galaxy and become clearly visible  only after  the image processing: a shape and nature of these inner arms may be different from  those observed in normal discy galaxies (See discussion in \citet{Lisker06}).

\begin{figure}
\hfill
\begin{minipage}[h]{0.47\linewidth}
\center{\includegraphics[width=1\linewidth]{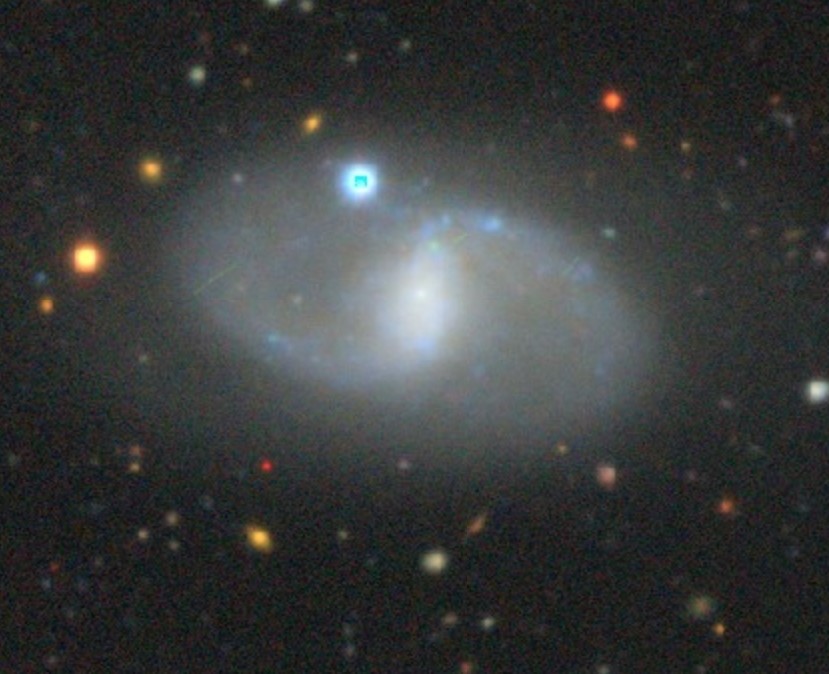}} \\ pgc 12961
\end{minipage}
\hfill
\begin{minipage}[h]{0.47\linewidth}
\center{\includegraphics[width=1\linewidth]{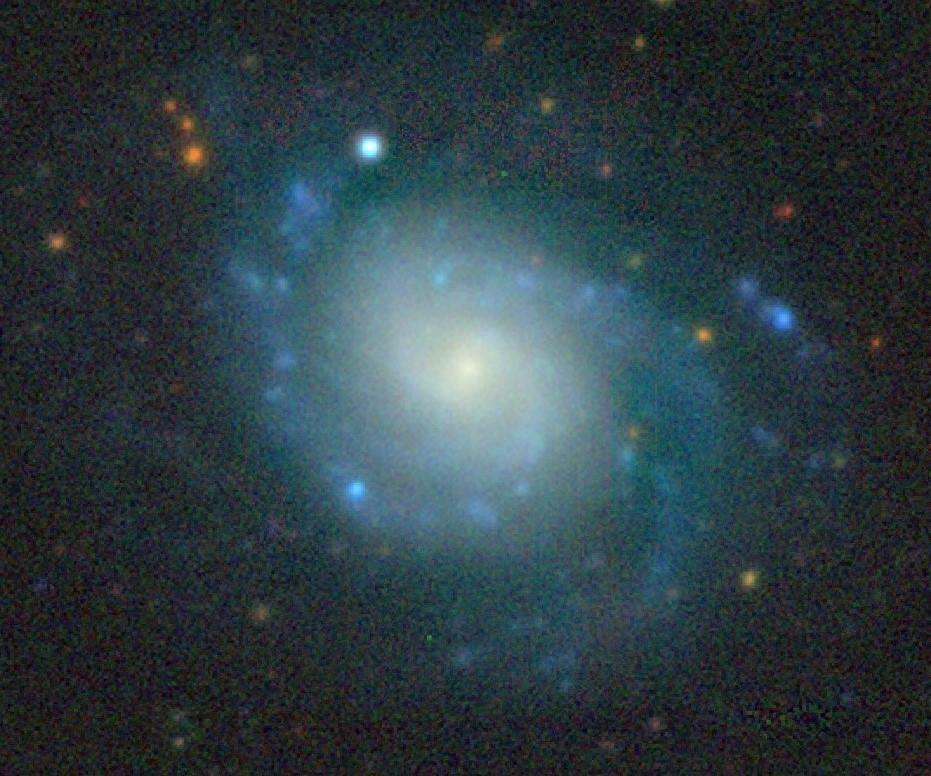}} \\pgc 49741
\end{minipage}

\vfill
\begin{minipage}[h]{0.47\linewidth}
\center{\includegraphics[width=1\linewidth]{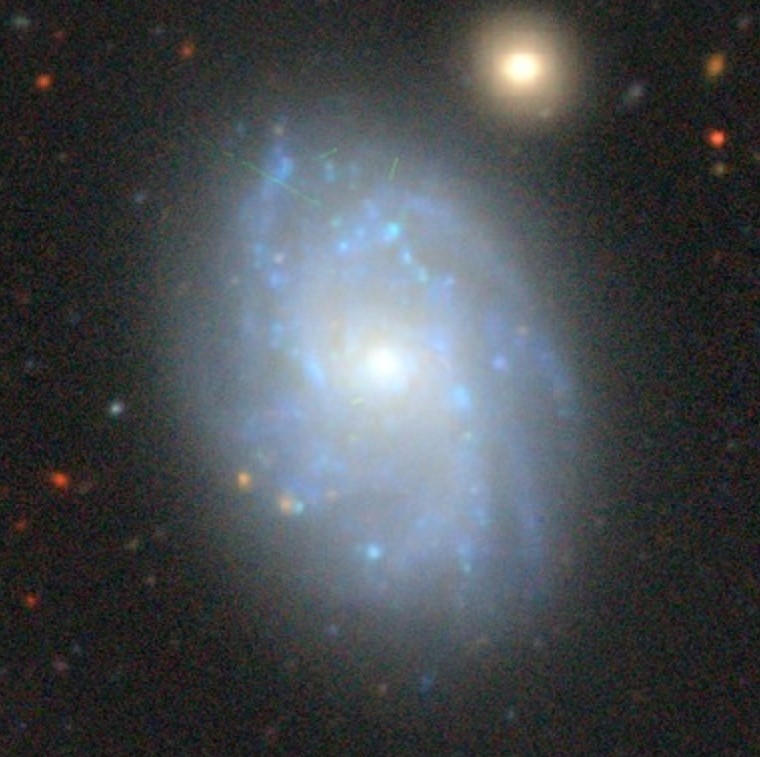}} pgc 12034 \\
\end{minipage}
\hfill
\begin{minipage}[h]{0.47\linewidth}
\center{\includegraphics[width=1\linewidth]{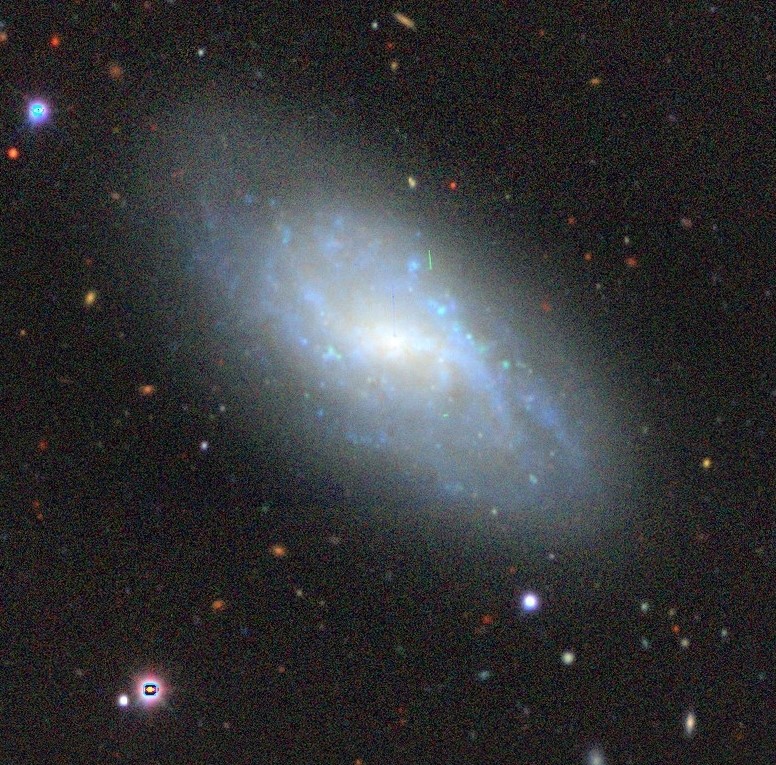}} pgc 41755 \\
\end{minipage}
\caption{The examples of dS galaxies. Up: dS with the robust Grand Design-type arm structure; bottom: dS galaxies with less ordered structure. Source: \href{http://legacysurvey.org}{Legacy Survey} images.}
\label{fig:examples}
\end{figure}

\section{The Sample of dwarf spiral galaxies}
\label{sec:sample}
In this work we refer to  late-type dwarf galaxies any low-luminosity  discy galaxy with an absolute magnitude $M_B>-18^m$ (following the \citet{Hidalgo04} sample), and a diameter of $D_{25}<12$~kpc (Hubble constant $H_0 = 75$\,km\,(sec\,Mpc)$^{-1}$). Diameter limitation is important in order to exclude objects whose low luminosity is a sequence of their low surface brightness ((LSB). Note that a galaxy having $M_B=-18^m, D_{25}<12$~kpc can not be considered as LSB galaxy. We limit the sample to nearby and sufficiently bright galaxies with not-too-high disc inclination in order to speak more confidently about their internal structure. For this purpose, we consider only galaxies with $m_B\lesssim 15^m$ and the inclination angle $i<75^\circ$. We also exclude from consideration nearby galaxies with a systemic velocity $V_{sys}< 500$\,km\,sec$^{-1}$ to avoid members of the local group and its environment.

As a first step, we selected from HyperLeda database all galaxies (324 objects) classified as spiral or irregular ones (Sa-Sm and Irr types, or numerical Hubble morphological class $T>0$), that satisfy the aforementioned restrictions. The vast majority of the  chosen galaxies belong to Sc--Irr types. Since many of the selected galaxies are relatively nearby objects for which the Hubble distance may be  not reliable, we check their distance estimates by comparing with those obtained from primary distance indicators in the Updated Nearby Galaxy Catalog (\citet{UNGC}). If at least one distance estimations cited in UNGC led to the luminosity value higher than $M_B=-18^m$, a galaxy excluded from consideration as possibly non-dwarf one. Eight excluded galaxies have  the PGC numbers 28390, 30531, 32767, 38167, 39423, 43927, 44672, 58501.  

Since the environment can affect the spiral structure formation and destruction, we also eliminate from the sample 5 members of the nearby clusters  (Virgo, UMa, and Fornax), parallel with 3 peculiar galaxies having clear signs of interaction with the close neighbors. However, we do not set a condition for galaxies to be isolated or single.

As the second step, we inspect visually the images of every selected dwarf galaxy to verify with some confidence the presence or the absence of a spiral-like or ring-like structures irrespective of how a given galaxy is classified in HyperLeda, thus forming the sample of 43 dwarf spiral galaxies. The rest of dwarf galaxies, where a large-scale spiral structure is not noticeable by visual inspection,  formed the comparison sample, containing 119 galaxies (all of them belong to types Sm and Irr).

Fig.~\ref{fig:examples} shows some examples of dS galaxies from our sample. The upper row presents two Grand Design-type galaxies: PGC 12961, which possesses  a strong bar (left) and the barless galaxy PGC 49741. Two galaxies at the bottom (PGC 12034 and PGC 41755)  demonstrate  blue flocculent arms, which makes them different from Irr galaxies.

The list of our sample of dS-galaxies and their main parameters are given in Table~\ref{tab:spiral} described below. In some cases a structure of galaxies we classify as dS is rather noisy due to non-homogeneous distribution of bright star-forming spots or arc-like filaments, so the presence of large-scale spiral arms  seems only the most likely. Galaxies which clearly demonstrate a Grand Design spiral pattern or (rare) a ring are separately marked by the asterisks.

\begin{table}\label{Spiral}
	 	\caption{\label{tab:spiral} The sample of dS-galaxies}
\begin{tabular}{lrrrrrrrr}
	PGC & $V_{sys}$ & $D_{25}$ & $V_{rot}$ & $\log M_{HI}$ & $\log M_{dyn}$ & $\log L_K$ & bar & $\log{(ii)}$ \\
	  & (km\,sec$^{-1}$) & (kpc) & (km\,sec$^{-1}$) & $(M_{\odot})$ & $(M_{\odot })$ & $(L_{\odot })$ &   &  \\
	\hline
	\hline
1292 & 1144 & 9.7 & 94 & 8.9 & 10.0 & 9.3 & - & 1.15\\
5981 & 1428 & 5.5 & 85 & 8.7 & - & 8.8 & - & 0\\
8602 & 1613 & 11.1 & 65 & 9.0 & 9.7 & 9.8 & B & 0.07\\
9539 & 2098 & 8.3 & 65 & 9.1 & 9.6 & 9.3 & B & -\\
9665 & 598 & 4.5 & 73 & 8.1 & 9.4 & 9.1 & - & 0.43\\
10586* & 584 & 3.8 & 61 & 8.4 & 9.2 & 8.8 & B & 0.43\\
12034 & 1495 & 8.4 & 107 & 8.6 & 10.0 & 9.2 & - & 0.07\\
12961* & 1672 & 8.5 & 117 & 8.4 & 10.1 & 9.5 & B & 0.17\\
12979 & 1529 & 8.2 & 97 & 9.1 & 10.0 & 9.3 & B & 0.07\\
14487 & 1856 & 9.5 & 41 & 9.0 & 9.3 & 8.6 & B & 0.55\\
14617 & 1368 & 9.0 & 91 & 8.6 & 9.9 & 9.3 & B & 0.14\\
14620 & 1277 & 8.4 & 57 & 8.1 & 9.5 & 9.8 & B & -\\
16617 & 1012 & 7.5 & 89 & 8.6 & 9.8 & 9.6 & B & 0\\
16864 & 973 & 9.7 & 121 & 9.0 & - & 8.8 & B & 0\\
19789* & 1441 & 8.6 & 167 & 9.4 & - & 9.5 & - & 0\\
23340 & 506 & 5.9 & 62 & 8.3 & 9.4 & 9.0 & B & 0\\
25427 & 1253 & 5.7 & 68 & 8.0 & 9.5 & 9.2 & - & 0.12\\
26231 & 1674 & 5.9 & 137 & 9.0 & 10.1 & 8.6 & B & 0.29\\
26465 & 1875 & 4.7 & 125 & 8.0 & - & 9.5 & - & 0\\
29549 & 1377 & 8.4 & 61 & 8.7 & - & 8.9 & B & -\\
30263 & 1297 & 2.8 & 113 & 8.4 & 9.6 & 9.3 & - & 0\\
31442 & 602 & 4.2 & 64 & 7.8 & 9.3 & 8.7 & - & 0\\
32287 & 1596 & 9.8 & 111 & 8.4 & - & 9.9 & B & 0\\
32390 & 1630 & 7.6 & 117 & 9.0 & 10.1 & 8.9 & B & 0.04\\
33813 & 1379 & 7.9 & 76 & 9.1 & 9.7 & 9.0 & - & 0\\
34156 & 1059 & 6.3 & 157 & 8.7 & - & 9.2 & - & 0\\
34466* & 1249 & 10.8 & 77 & 8.4 & 9.9 & 9.0 & B & 0\\
36137* & 1145 & 9.0 & 66 & 8.8 & 9.7 & 9.4 & - & 0.22\\
36305 & 1449 & 6.0 & 66 & 8.4 & 9.5 & 8.8 & B & 0\\
37132* & 921 & 9.6 & 97 & 8.9 & 10.0 & 8.8 & B & 0.24\\
37550 & 886 & 4.1 & 130 & 8.4 & - & 8.7 & - & 0\\
37853 & 1569 & 7.8 & 176 & 9.0 & - & 9.4 & B & 0.05\\
39237 & 700 & 2.8 & 72 & 8.2 & 9.2 & 8.8 & - & 0.08\\
39837 & 1099 & 3.3 & 21 & 7.9 & 8.2 & 8.1 & B & 0.07\\
41755 & 1175 & 9.1 & 74 & 8.6 & 9.8 & 9.5 & B & 0.03\\
44089 & 709 & 4.4 & 65 & 8.4 & 9.3 & 8.7 & - & 0.32\\
45728 & 1064 & 5.7 & 35 & 8.5 & 8.9 & 8.4 & B & 0.14\\
45858 & 792 & 2.7 & 76 & 8.0 & 9.3 & 8.7 & B & 0\\
47577 & 1017 & 4.7 & 60 & 8.2 & 9.3 & 9.1 & - & 0.72\\
49741* & 1619 & 8.8 & 95 & 8.8 & 10.0 & 8.9 & - & 0.01\\
51655 & 1148 & 7.0 & 52 & 8.4 & 9.4 & 8.6 & B & 1.51\\
51971* & 1528 & 4.0 & 111 & 8.7 & 9.8 & 8.6 & B & 0.21\\
57678 & 717 & 7.1 & 54 & 8.5 & 9.4 & 8.9 & B & 0\\
	\hline 
\end{tabular}

\textbf{Notes}: \textbf{PGC} is the name of a galaxy; \textbf{$V_{sys}$} is systemic velocity; \textbf{$D_{25}$} is the linear size; \textbf{$V_{rot}$} is maximum rotation velocity; \textbf{log$M_{HI}$} is logarithm of the HI mass; \textbf{log$M_{dyn}$} is the logarithm of the dynamical mass; \textbf{log$L_K$} is the logarithm of the galaxy luminosity in K-band; \textbf{bar}  marks the presence or absence of a bar; \textbf{log$(ii)$} is the logarithm of isolation index (see the text). Dwarfs with Grand Design structure are marked with asterisks (\textbf{*}) in the first column.
\end{table}
\clearpage

 Table~\ref{tab:spiral}  presents some parameters of chosen galaxies, most of them were taken on the basis of HyperLeda data: heliocentric systemic velocity $V_{sys}$; galaxy diameter $D_{25}$ at the isophotal level 25 mag/arcsec$^2$ in the B-band; rotation velocity $V_{rot}$, which in most cases was estimated from the HI line width  after correction for a disc inclination (see $V_{rot}$ from HyperLeda database); HI mass calculated from HI-line flux after correction for self-absorption (see the parameter $m21c$ in HyperLeda database); the presence or absence of a bar. Table also contains luminosities $L_K$ in solar units, obtained from the K-magnitudes of 2MASS catalog (\citet{2MASS} (the solar absolute magnitude was taken  as $M_{\odot }^K = 3.27$ (\citet{Cohen03})), and the dynamic mass of the galaxy within $D_{25}$ determined by the simplified relation:
\begin{equation}\label{eq-Mdyn}
M_{dyn}=\frac{V_{rot}^2D_{25}}{2G} \,.
\end{equation}

In this work we use $V_{rot}$ and related quantities only for those galaxies where the angle $i$ of  disc inclination  exceeds $35^\circ$ (according to HyperLeda),  otherwise  the  estimates of $V_{rot}$ appear unreliable. 

The last column indicates the logarithm of isolation index $\log(ii)$ taken from the full electronic version of the Catalog of the Nearby Galaxy Groups (\citet{MK_groups}). The cited  authors proposed a clusterization algorithm  which they used for selection of gravitationally bound pairs (\citet{Karach&Makarov08}),  groups (\citet{Karachentsev95, MK_groups}) and isolated galaxies in the Local Volume  (\citet{Karachentsev_etal11}). They considered every virtual pair of a given galaxy with the nearest neighbour   as a virtually bound one, assuming that for a given pair to be bound  a module of  potential energy of interaction between galaxies should be larger than their total kinetic energy. To check its possibility,  they calculated for each virtual pair of galaxies the module of gravitational-to-kinetic energy ratio  $V_{ik}^2R_{ik}/2GM_{ik}$,  where $V_{ik}$  and $R_{ik}$ are the radial velocity differences and the projected separations of the virtual pair components, $M_{ik}$ is their total mass, expressed in terms of the K-luminosity assuming $M_{ik}/L_K$=\,6. 
The isolation index $(ii)$  they calculated  shows  how many times it is necessary to increase the $L_K$-based estimated mass of galaxies in a given pair in order this pair to be  gravitationally bound  (see f.e. \citet{Karachentsev_etal11}).  For example, $\log(ii)\approx0$ means that a given galaxy is most probably a member of gravitationally bound pair or  a group. Although the mass estimates are subject to projection effects and have only a statistical meaning, a comparison of the isolation indices of galaxies of different samples allows us to compare them by the expected  gravitational influence from their neighbours which galaxies of the samples may experience. 

Surprisingly, there are only 6 dS-galaxies, common with the sample of \citet{Hidalgo04}. The reason is that we use different source material (HyperLeda vs UGC catalog) and different selection criteria (our restrictions are tougher with the exception of the a little higher diameter limit). An important role is played by the fact that a significant fraction of galaxies of Hidalgo-G{\'a}mez’ sample are fainter than  $B = 15^m$ according to HyperLeda, so they were not included in our list.

\section{Comparison of dwarf spiral and non-spiral galaxies}
\label{sec:statistic}
\begin{figure}
	\includegraphics[width=\columnwidth]{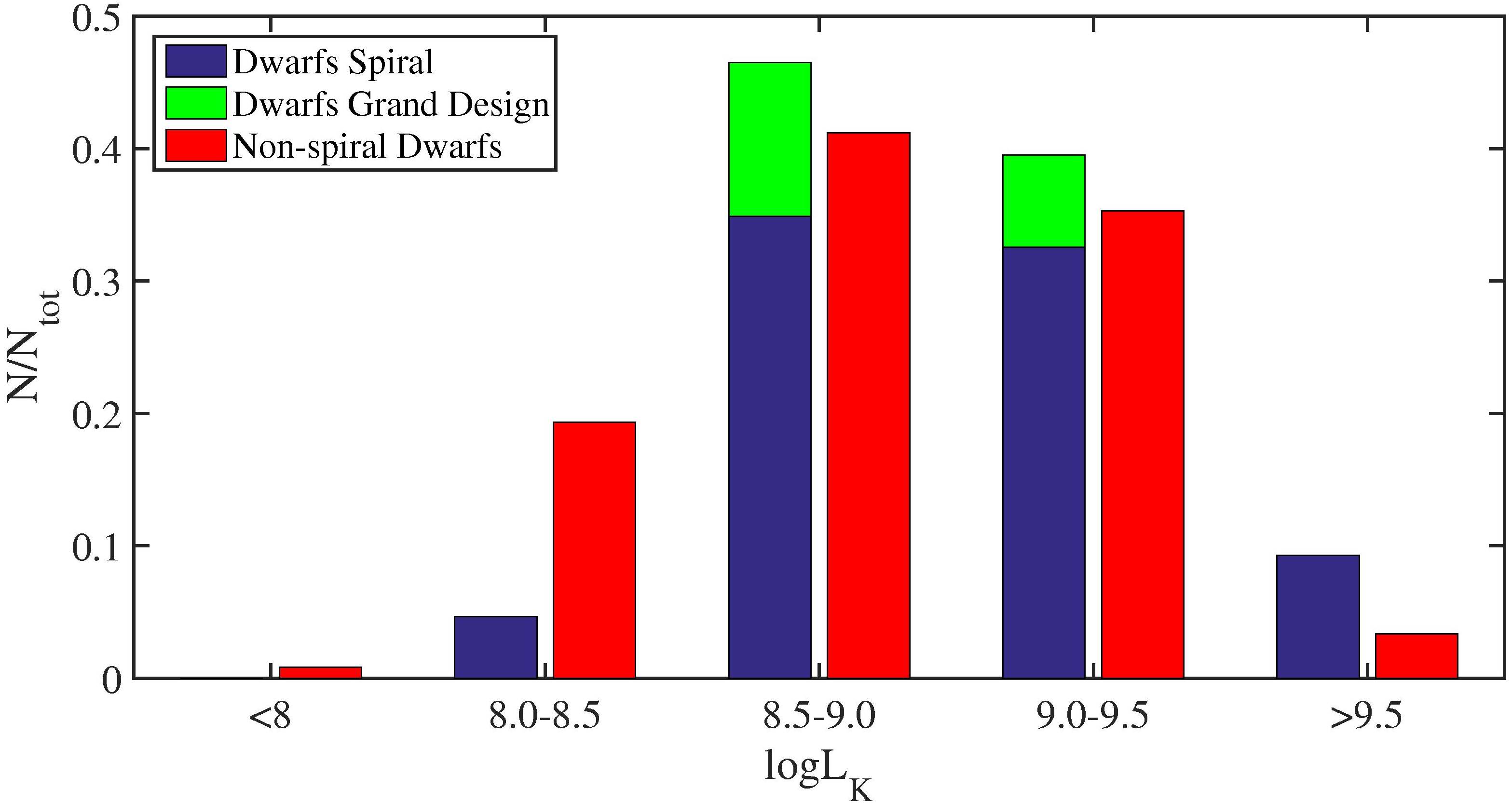}
	\caption{A normalized number of dS and non-spiral  galaxies for five luminosity intervals. Green color marks Grand Design dS-galaxies.}
	\label{fig:hist_LK}
\end{figure}

Fig.~\ref{fig:hist_LK} presents the K-band luminosity distributions of dS and non-dS (irregular) galaxies. K-band was chosen because NIR luminosity better describes stellar mass of a galaxy. The steep decline in the number of galaxies with $\log L_K > 9.5$ simply reflects the restrictions that we have imposed on the luminosity $L_B$ of galaxies when sampling. The histogram confirms the conclusion of \citet{Hidalgo04}, that dS-galaxies are mostly found among the most massive dwarf systems. Indeed, among galaxies fainter than $\log L_K = 8.5$, there are only two dS-galaxies (several percent of our sample) and about 20~percent  galaxies of dIrr sample.

\begin{figure}
	\includegraphics[width=\columnwidth]{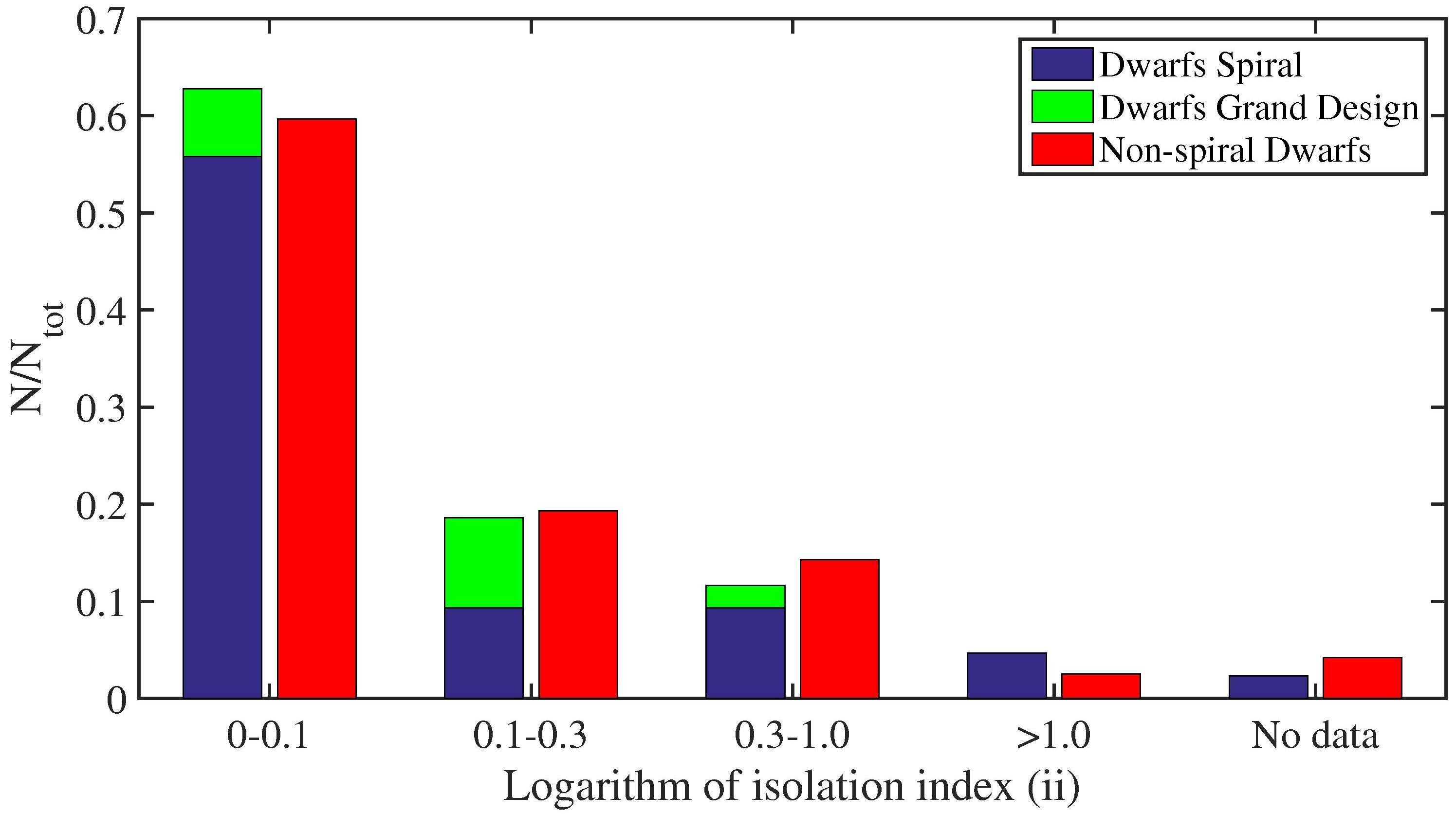}
	\caption{A normalized number of dS and non-spiral  galaxies for five intervals of isolation indices (\textit{ii}). Green color marks Grand Design dS-galaxies. Galaxies with higher values of (\textit{ii}) are more isolated}.
	\label{fig:hist_II}
\end{figure}

Distribution of isolation indices  (see Fig.~\ref{fig:hist_II}) shows that although both dS and non-dS galaxies can in no way be considered as highly isolated, their index distributions are similar. Grand Design galaxies (marked by green) also do not reveal a tendency to have low values of the index. It gives evidence that the external influence on a galaxy can hardly play the essential role in the formation of a spiral structure in low-mass systems.

\begin{figure}
	\includegraphics[width=\columnwidth]{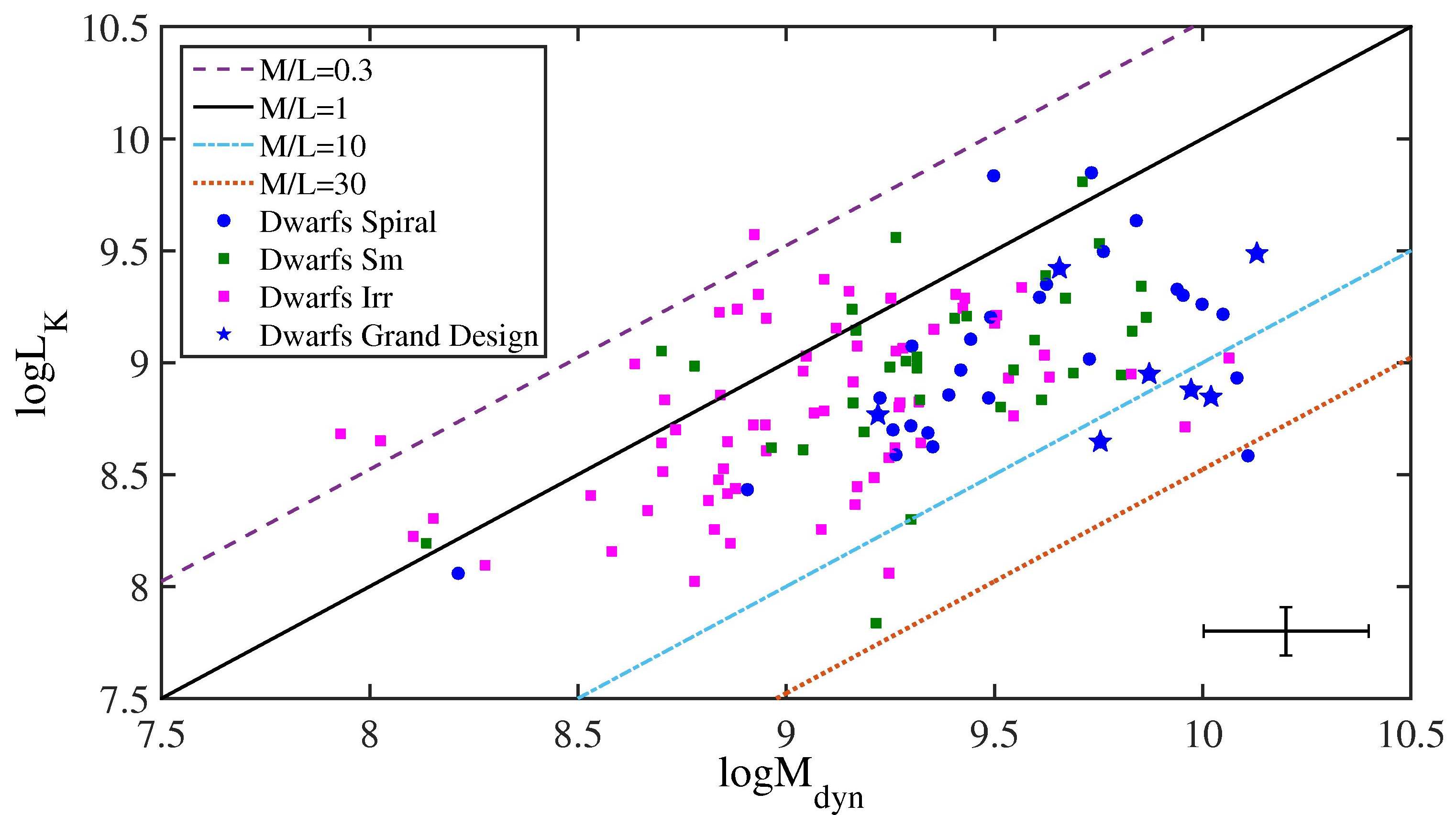}
	\caption{Luminosity $L_K$ vs $M_{dyn}$  for dwarf galaxies. dS-galaxies are marked by blue symbols. Straight lines correspond to different $M_{dyn}/L_K$ ratios. Most dwarf galaxies lay between  $M_{dyn}/L_K$= 1 and 10 solar units regardless of the presence of spiral structure. Error bars present root mean square errors averaged over the sample.}
	\label{fig:L_K}
\end{figure}

\begin{figure*}
\hfill
\begin{minipage}[p]{0.49\linewidth}
\centering \subfigure[]{
\includegraphics[width=1\linewidth]{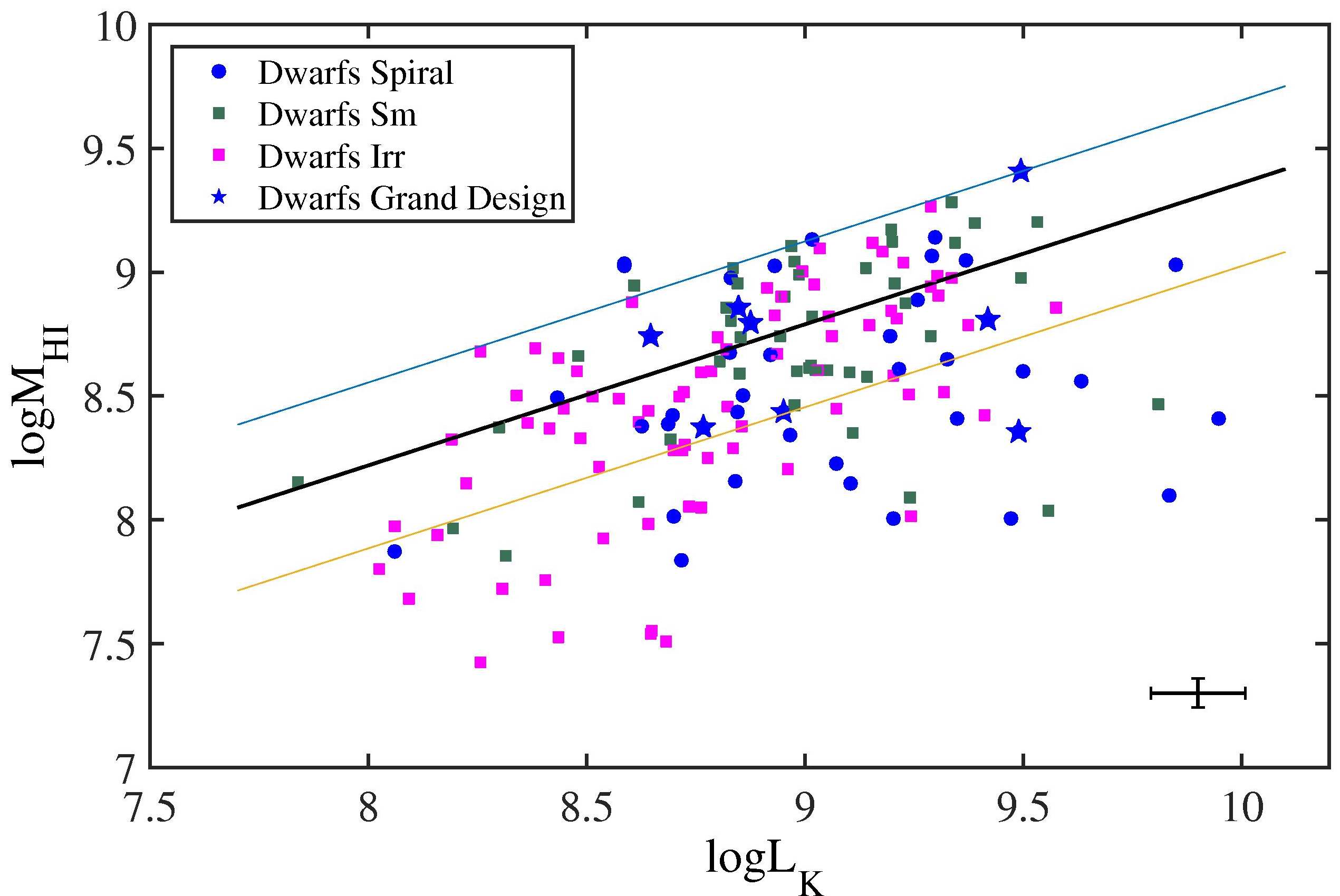}
\label{fig:LK_MHI}}
\end{minipage}
\hfill
\begin{minipage}[p]{0.49\linewidth}
\centering \subfigure[]{
\includegraphics[width=1\linewidth]{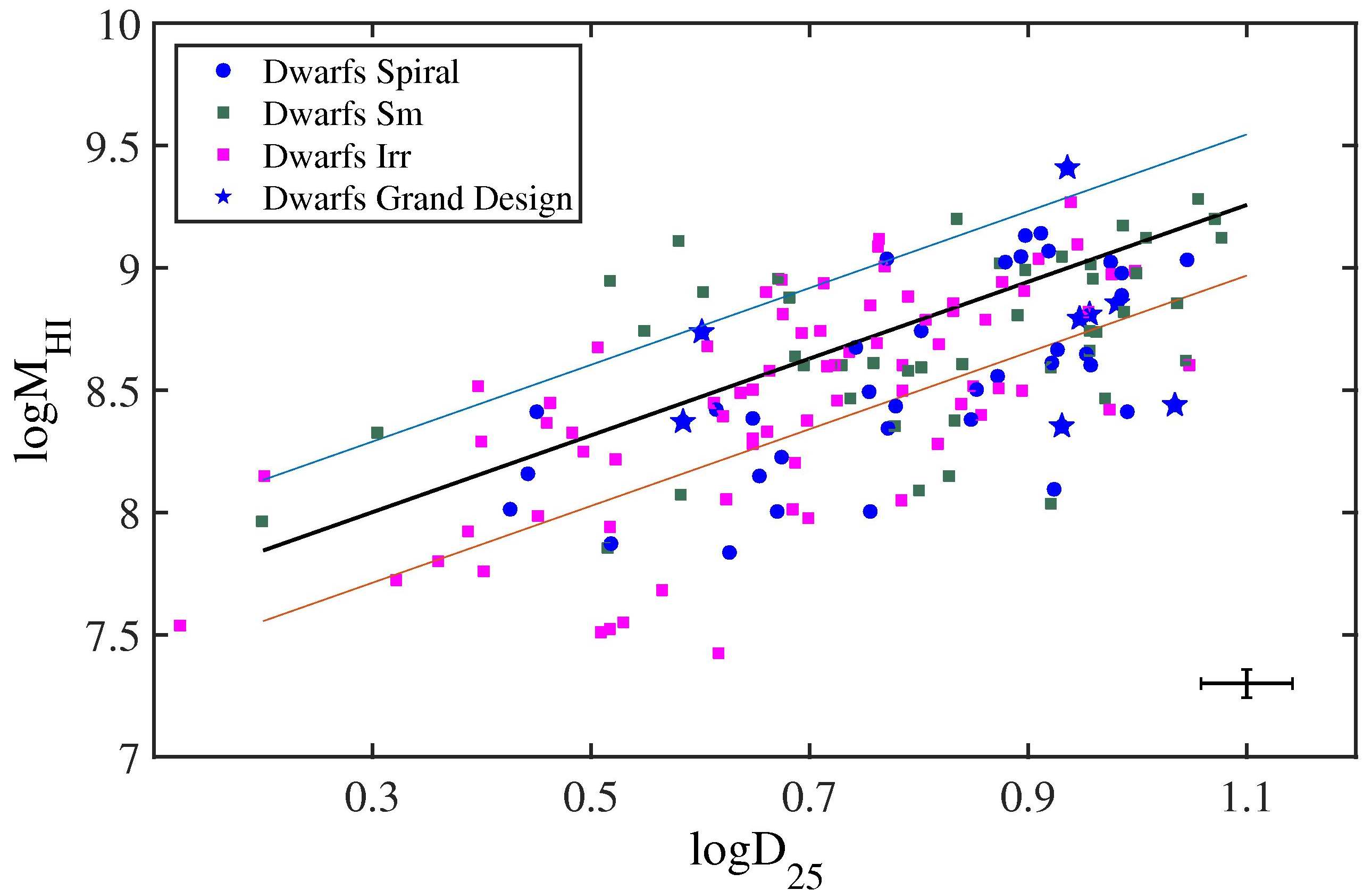}
\label{fig:MHI_D}}
\end{minipage}
\vfill
\begin{minipage}[p]{0.49\linewidth}
\centering \subfigure[]{
\includegraphics[width=1\linewidth]{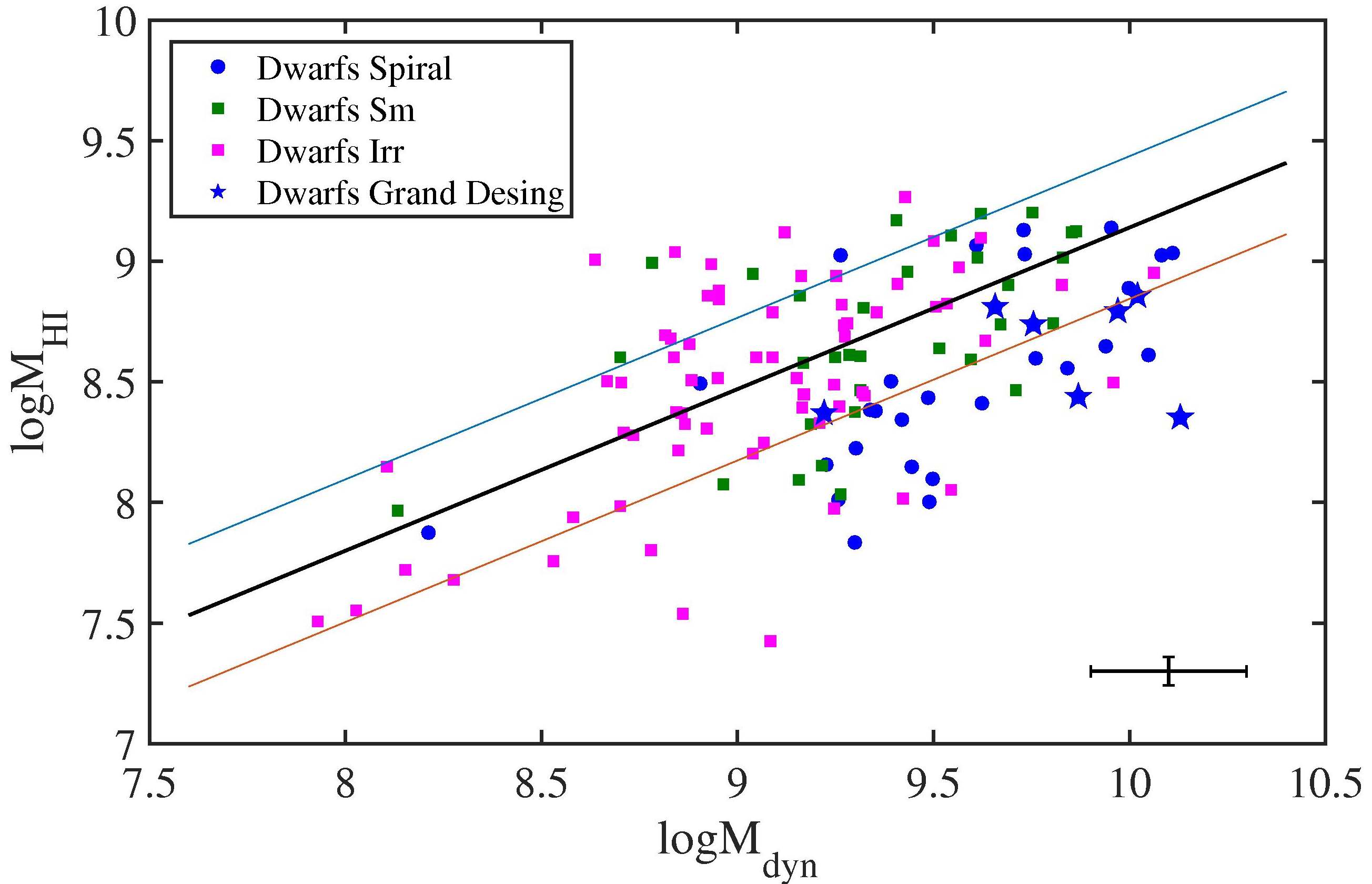}
\label{fig:MHI_Mdyn}}
\end{minipage}
\hfill
\begin{minipage}[p]{0.49\linewidth}
\centering \subfigure[]{
\includegraphics[width=1\linewidth]{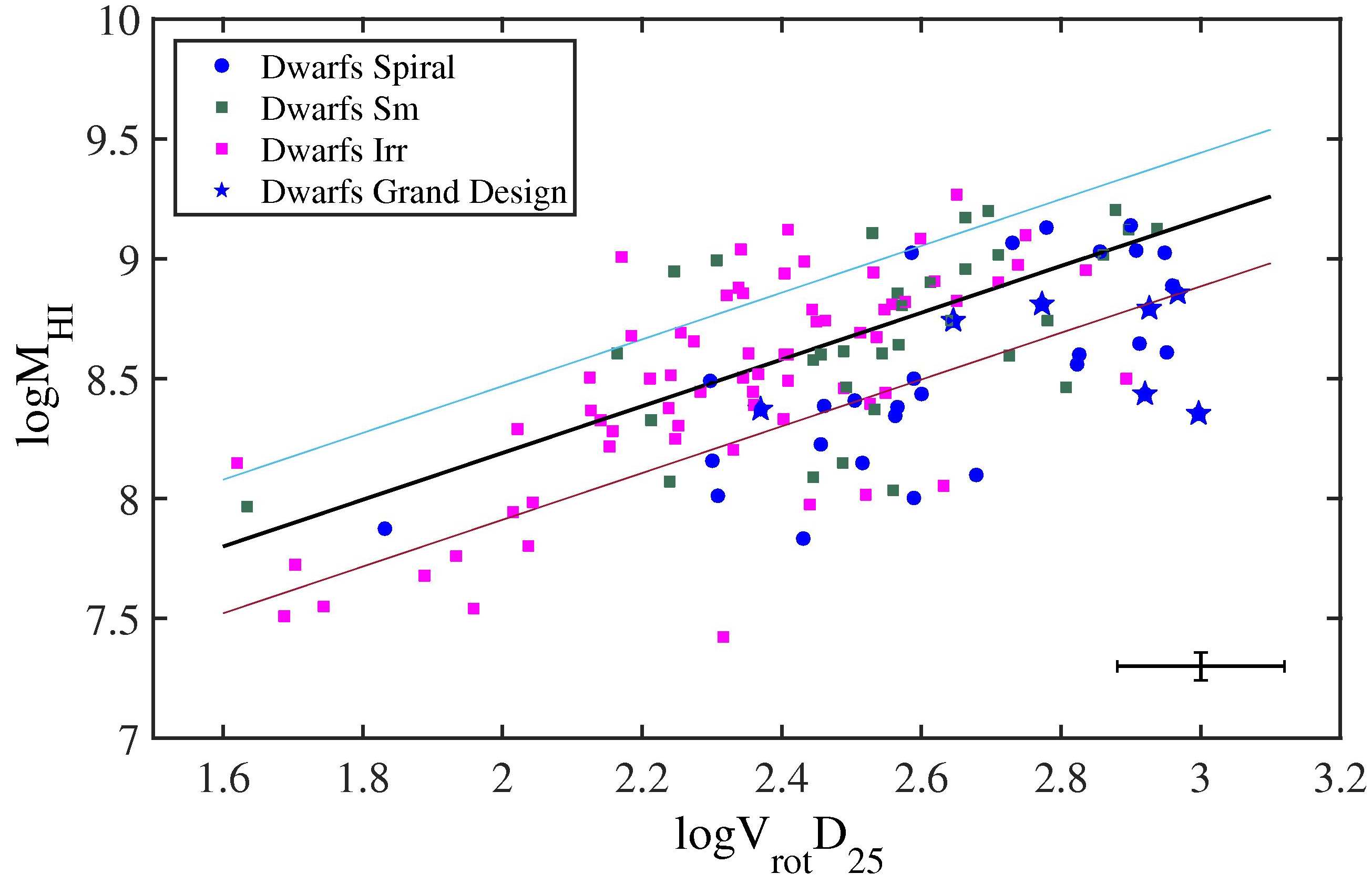}
\label{fig:MHI_VD}}
\end{minipage}
\vfill
\begin{minipage}[p]{0.49\linewidth}
\centering \subfigure[]{
\includegraphics[width=1\linewidth]{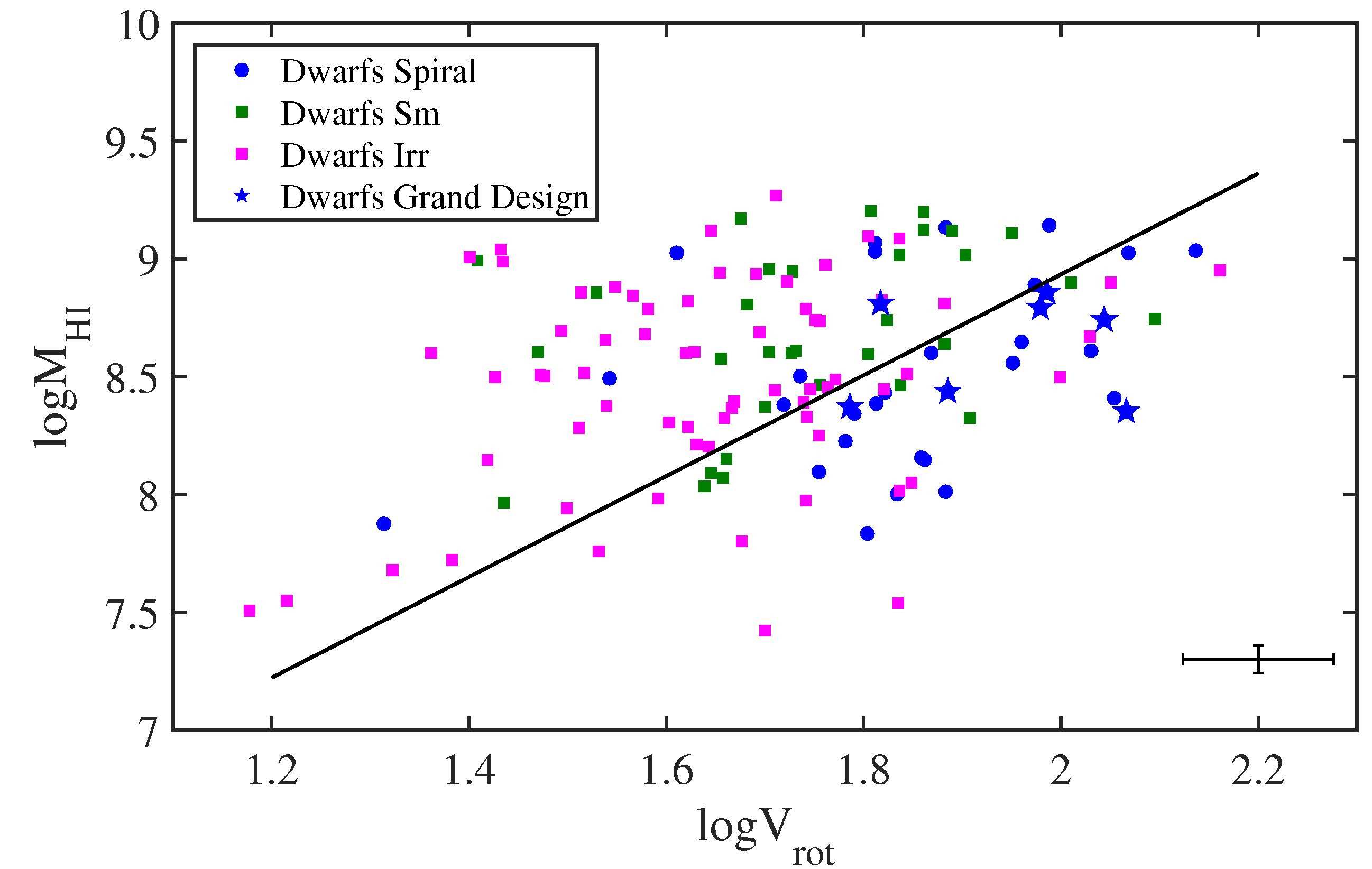}
\label{fig:MHI_V}}
\end{minipage}
\caption{The relations between different parameters of dwarf galaxies. Symbols are the same as in Figure~\ref{fig:L_K}. For panels \subref{fig:LK_MHI}-\subref{fig:MHI_VD} the straight black  line is the extrapolated regression line obtained for {isolated late-type spiral galaxies of AMIGA survey} (see the text). Two color lines on both sides from the regression line illustrate a dispersion ($1\sigma$) for isolated galaxies. For panel \subref{fig:MHI_V} the straight line marks the relationship taken from \citet{Local_TF17} for dwarf galaxies. Error bars present root mean square errors averaged over the sample.}
\label{dwarf_relations}
\end{figure*}

\begin{table}
	 	\caption{\label{tab:dev_AMIGA} Mean deviations of dwarf galaxies from the extrapolated regression lines for AMIGA galaxies (\citet{Zasov&Zaitseva17}), and (the last line) from the Tully-Fisher relation for local dwarfs  \citet{Local_TF17}.}
\begin{tabular}{lccc}
	& dS & dSm & dIrr  \\
	\hline
	$M_{HI} - L_K$ & -0.31$\pm$0.07 & -0.06$\pm$0.06 & -0.15$\pm$0.04 \\
	$M_{HI} - D_{25}$ & -0.25$\pm$0.05 & -0.09$\pm$0.06 & 0.05$\pm$0.04 \\
	$M_{HI} - M_{dyn}$ & -0.38$\pm$0.06 & 0.01$\pm$0.06 & 0.07$\pm$0.06 \\
	$M_{HI} - V_{rot}D_{25}$ & -0.34$\pm$0.05 & -0.01$\pm$0.06 & 0.06$\pm$0.05 \\
	$M_{HI} - V_{rot}$ & -0.16$\pm$0.07 & 0.35$\pm$0.08 & 0.39$\pm$0.08 \\
	\hline 
\end{tabular}
\end{table}

Fig.~\ref{fig:L_K} compares luminosity $L_K$ and dynamical mass $M_{dyn}$ of the samples of dS (blue) and non-dS galaxies of Sm and Irr types.  Parallel lines correspond to the different dynamic  mass-to-light ratios $M_{dyn}/L_K$= 0.3, 1, 10 and 30 solar units. This ratio depends primarily on the relative fraction of the dark mass within the optical radius, and to a lesser extent, on the characteristics of  stellar population and gas content. In the diagram, the $ M_{dyn} / L_K$ ratios have a large spread. Although partially it may be caused by the line-width-based mass estimates errors, it agrees with a high dispersion of mass-to-light ratio of dwarf galaxies found by \citet{McCall_etal_2012}. Dwarf spiral  galaxies, marked by blue symbols are well mixed among the non-spiral dwarfs with similar values of $M_{dyn}$.  The error bars in the right lower corner of this and the next diagrams give the characteristic values of root mean square errors of the compared parameters.

Note that a general position of dwarf galaxies at the diagram demonstrates that most of them are characterized by high ratios $M_{dyn}/L_K>1$  regardless of  their internal structure. If to take into account that for stellar population of star-forming galaxies (including the dwarf ones) the expected  ratio $M_* / L_K$ does not exceed 0.7 solar units (See f.e. \citet{Ponomareva18}), a significant contribution of dark halo to the total mass of dwarf galaxies within their optical radii is obvious. 

The next five diagrams compare the  mass of neutral hydrogen $M_{HI}$ in dS and non-dS galaxies, using the relations between total mass of HI and the total luminosity [$L_\odot$] (Fig.~\ref{fig:LK_MHI}), optical diameter [kpc] (Fig.~\ref{fig:MHI_D}), dynamical mass [$M_\odot$] (Fig.~\ref{fig:MHI_Mdyn}), specific angular momentum [km/sec $\cdot$ kpc] (Fig.~\ref{fig:MHI_VD}), and $V_{rot}$ [km/s] (Fig.~\ref{fig:MHI_V}). In  Figs.~\ref{fig:MHI_D}--\ref{fig:MHI_V} we limited ourselves to galaxies with $L_K\geq5\cdot10^8L_\odot$, since among the  objects of lower luminosity there are practically no dS- galaxies (See Fig.~\ref{fig:LK_MHI}). 

To take into account a general tendency  of $M_{HI}$ to vary parallel with the other parameters of galaxies,  in every diagram we insert a reference sequence,  with respect to which one can compare $M_{HI}$  for galaxies of different samples. In the first four diagrams we use as the reference line the relationship obtained for late-type isolated galaxies of AMIGA sample (\citet{AMIGA07}) with an inclination $i>35^\circ$ constructed on the basis of Karachentseva’s Catalogue of Isolated Galaxies (\citet{CIG}) (see Table~3 in \citet{Zasov&Zaitseva17}). AMIGA galaxies are a fairly convenient comparison sample, as they are late type spirals with no explicit signs of interaction with the other galaxies. Since the majority of AMIGA galaxies are not dwarfs by their luminosity, here we  use the linear extrapolations of the AMIGA sequences to the low sizes and rotation velocities. In the figures, these extrapolated relationships are shown by thick black lines. Two parallel lines on both sides illustrate a points spread ($1 \sigma$) for AMIGA galaxies. 
For comparison of dS-galaxies and dwarf galaxies at the diagram  $M_{HI} - V_{rot}$ we take as the reference line the  relationship between the velocity of rotation  found from HI line-width and the mass of HI given by \citet{Local_TF17} for local volume galaxies of the  Updated Nearby Galaxy Catalog  (a straight line in Fig.\,\ref{fig:MHI_V}). 

Table~\ref{tab:dev_AMIGA} gives the mean deviations $\Delta M_{HI}$ for galaxies of our samples from the regression lines of the reference sequences at the diagrams given above (see the legend of Table~\ref{tab:dev_AMIGA}). The mean difference $\Delta M_{HI}$ is $0.33\pm 0.06$ dex between dS and dSm galaxies and  $0.37\pm 0.07$ dex between dS and dIrr galaxies.  
The relationship $M_{HI}$ vs $V_{rot}$  (see diagram in Fig.\,\ref{fig:MHI_V}) shows a low mean  deviation of dS-galaxies from the regression line $\Delta M_{HI} = 0.16\pm 0.04$, although if one compares dS with non-dS-galaxies, the difference between them is  slightly higher:  $\Delta M_{HI}\approx 0.5$. However the latter value is very approximate   due to a large spread of non-spiral dwarfs at  the diagram. 

To conclude, a general result of these comparisons is that the mass of hydrogen in dS-galaxies is, on average, at $\sim$0.35~dex, or approximately two times lower than in  dwarf non-spiral  galaxies with similar dynamic and photometric characteristics. However, it should be borne in mind that most dS-galaxies are nevertheless rather  rich of gas, if to compare  their mass of HI with  stellar mass or luminosity  (See Table~\ref{tab:spiral}).
 
We also find that a significant part of galaxies we consider have a bar, regardless of the presence of a spiral structure. It evidences that the existence of a bar does not play a decisive role for the development of spiral arms. The very fact that many dwarf galaxies have a bar allows to suggest that  bars are  long-lived structures in such slowly rotating systems as dwarf galaxies.
A favorable condition for the formation of bars in dwarf galaxies most probably is associated with a lower concentration of matter towards the center due to the absence of massive bulges.

\section{Numerical simulation of spiral arms formation in dwarf galaxies}
\label{sec:options}

\subsection{Formulation of the problem}
\label{subsec:NumMode1}
As it was shown above, the presence or the absence of spiral arms in dwarf galaxies is not explicitly associated with the environment or with the formation of well developed  bar. Therefore, it makes sense to see conditions  of spiral arm  formation in low mass galaxies as a result of internal processes in their discs. A link between disc parameters and the  development of spiral-forming instability can be investigated using the numerical simulations of dynamical evolution of stellar/gaseous discs. 

We proceed from the natural assumption that the existence of spiral structure is associated with a self-gravity of a disc. Since spirals in low-luminosity galaxies are observed only in a small percentage of cases, it is important to understand the conditions under which a formation of a large-scale density waves  is possible.

In the models described below we try to find the parameters of slowly rotating discs which lead to the development of a  long-lived global spiral n pattern. We used 3-component  models of bulgeless galaxies, consisting of stellar and gaseous discs  immersed in a rigid massive spherical pseudo-isothermic component (a dark halo), where maximum rotational velocity of gas (which is close to a circular velocity) $V_g^{max}$ does not exceed 100~km\,sec$^{-1}$.

{The conditions for the formation and survival of a bar are not considered in this paper. This is a special topic, which is devoted to many works  (See f.e. \cite{Polyachenko2020bar} and the references therein). Here we proceed from the fact that  bars are observed in a significant fraction of dwarf galaxies, regardless of whether they have a spiral structure or not.}

For all models we assume that initially the turbulent velocity dispersion of gas $c_g$ was equal to  $\approx 10$~km\,sec$^{-1}$ allover a disc. It agrees with observations of both dwarf and non-dwarf galaxies (See f.e. \citet{Stilp-etal-2013kinem}), although in real spiral galaxies the velocity dispersion of gas is usually slightly lower in the outer regions of HI disc (See for example \citet{Tamburo_et_al09, Ianjamasimanana15}. During the numerical evolution of model galaxies the values of  $c_g$ may change locally, however the equation of state of gas we use is  close to the isothermal one, so the local variations of $c_g$ remain small. 

To limit our models to  dwarf galaxies, we accept the ratio of maximal circular velocity of a disc  to a gas velocity dispersion  $V^{\max}_g/c_g\le 10$, and mass of gas  vs stellar mass ratio ${M}_g/M \le 1$. Velocity of rotation of gas at the disc periphery was restricted by the interval  $V^{\max}_g \simeq 30\div 100$\,km\,sec$^{-1}$ (we take into account that for the smallest values of $V^{\max}_g$ its value is slightly lower than the circular velocity). At the same time we assume that the vertical velocity dispersion of stars $c_z$ is higher than the turbulent velocity of gas $c_g \approx 10$~km\,sec$^{-1}$ everywhere.  For slowly rotating galaxies it means that   their discs may be  thick and dynamically hot to compare with giant galaxies, especially in the outer regions, where a ``vertical'' gradient of gravitational potential is shallow.

 We  also take into account that the stellar velocity dispersion is anisotropic in the radial ($c_r$), azimuthal ($c_\varphi$) and vertical ($c_z$) direction. It was assumed that  $c_\varphi = c_r \varkappa / 2\Omega$ (the Linblad relation), and $c_z \simeq (0.5\div 0.8)\cdot c_r$ (\citet{Pinna_et_al18}), \citet{Shapiro_et_al03}).  From this it follows, that  $c_\varphi \simeq c_z < c_r$.
 
\subsection{Numerical implementation of models}
\label{subsec:NumMode2}
Dynamics of a collisionless stellar disc consisting  of $N_*$  equal point masses $m_i$ is determined by a system of equations ($i=1,2, \dots, N_*$)
\begin{equation}\label{eq-Nbody}
\frac{d^2 {\vec r}_{i}}{dt^2} = \sum\limits_{j=1,\, j\ne i}^{N_*}  \vec{f}^{(s)}_{ij} + \vec{F}_i^{(gas)} + \vec{F}_i^{(ext)} \,,
\end{equation}
where ${\vec r}_{i}$  is the radius-vector of $i$-th particle, 
 $\vec{f}_{ij}$  is the interaction force between $i$-th and $j$-th particles, $\vec{F}_i^{(gas)}$ and $\vec{F}_i^{(ext)}$ are the gravitational forces acting on the particle from  gaseous disc and spherical components of galaxy respectively. 
 
 The gaseous component is described by the standard equations of hydrodynamics, which we write in a form convenient for applying the SPH-method ($i=1, \ldots, N_g $):
\begin{equation}\label{eq-hydrodynamics-U}
\frac{d \vec{U}_i}{d t} = - \frac{1}{\varrho_{gi}} \nabla \, p_i + \sum\limits_{j=1,\, j\ne i}^{N_g}  \vec{f}^{(g)}_{ij} + \vec{F}_i^{(star)} + \vec{F}_i^{(ext)}  \,,
\end{equation}
\begin{equation}\label{eq-hydrodynamics-E}
\frac{d \varepsilon_i}{d t} = - \frac{p_i}{\varrho_{gi}}\,\nabla \vec{U}_i    \,,
\end{equation}
where $\varrho_{g}$ is the gas volume density, $\vec{U}$ is the gas velocity vector, $p$ is the  pressure, $\vec{F}^{(star)}$ is gravitational force associated with  stellar component, $\varepsilon$ is the internal energy. We use the equation of state for ideal gas in the form: 
$$
\varepsilon_i = \frac{p_i}{(\gamma - 1)\,\varrho_{gi}} \,,
$$
where a choice of the adiabatic index $\gamma$ ensures the quasi-isothermal state of gas. 

{Our numerical models are based on the direct N-body simulations for  stellar disc component and on the Lagrangian approach using the SPH-method for the gas component. The parallel CUDA-realization of numerical SPH-code was described in detail by~\citet{A2017_SPH} for CPU+GPU computing systems.  A description of how the parallel CUDA-algorithm  may be used for N-body modeling is illustrated in~\citet{A2018_Nbody}. The gravitational force is calculated by the direct Particle-Particle  method of summing the gravitational interaction of each particle with all the other stellar and gaseous particles:
\begin{equation}\label{eq-gravForcePP}
 \vec{f}_{ij} = -G\,\frac{m_j\,\vec{r}_{ij}}{(r_{ij}^2 + r_c^2)^{3/2}} \,, 
\end{equation}
where $r_{ij}$ is a distance between two particles, $r_c$ is a cut-off radius, which ensures the collisionless for the stellar component and reduce a strong pairwise gravitational interactions of close gas particles. A self-consistent modeling of dynamics of collisionless and gaseous particles  using a parallel CUDS algorithm ``N-body+SPH'' was described by~\citet{Khrapov19}. The application of this code for dynamical modeling of stellar/gaseous disc of the Milky Way may be found in~\citet{A2020dracones}.}

We use the equal number $N$ of particles with masses $m_*$ and $m_g$, representing stars and gas respectively. Thus,  for a total mass of stellar and gas components we have  $M_*=m_*N$ and  $M_g=m_gN$. A series of test numerical experiments with different total number of particles $N_{sum}=2N$ in the range $2^{19} \div 2^{23}$ showed a satisfactory convergence of solutions for $N_{sum} \ge  2^{21}\simeq 4\cdot 10^6$. All the models described below satisfy this condition.

Note that for  numerical simulation of evolution of gravitationally unstable discs it is important to start with the equilibrium state for both gaseous and stellar components. Therefore, to construct the initial quasi-stationary configuration of a polytropic gaseous disc embedded in the total gravitational potential of a galaxy, we use the condition of hydrostatic equilibrium in the radial and vertical directions
$$
- \frac{V_g^2}{r} = - \frac{d p}{\varrho d r} + \vec{F}_r^{(ext)} + \vec{F}_r^{(star)} + \vec{F}_r^{(gas)} \,, 
$$
$$
\frac{d p}{\varrho d z} = \vec{F}_z^{(ext)} + \vec{F}_z^{(star)} + \vec{F}_z^{(gas)} \,. 
$$
The equilibrium structure of stellar disc with a scale height $z_0(r)$ was achieved using the iteration procedure (See \cite{khoperskov-bizyaev2010} for details).

\begin{figure}
	\includegraphics[width=\columnwidth]{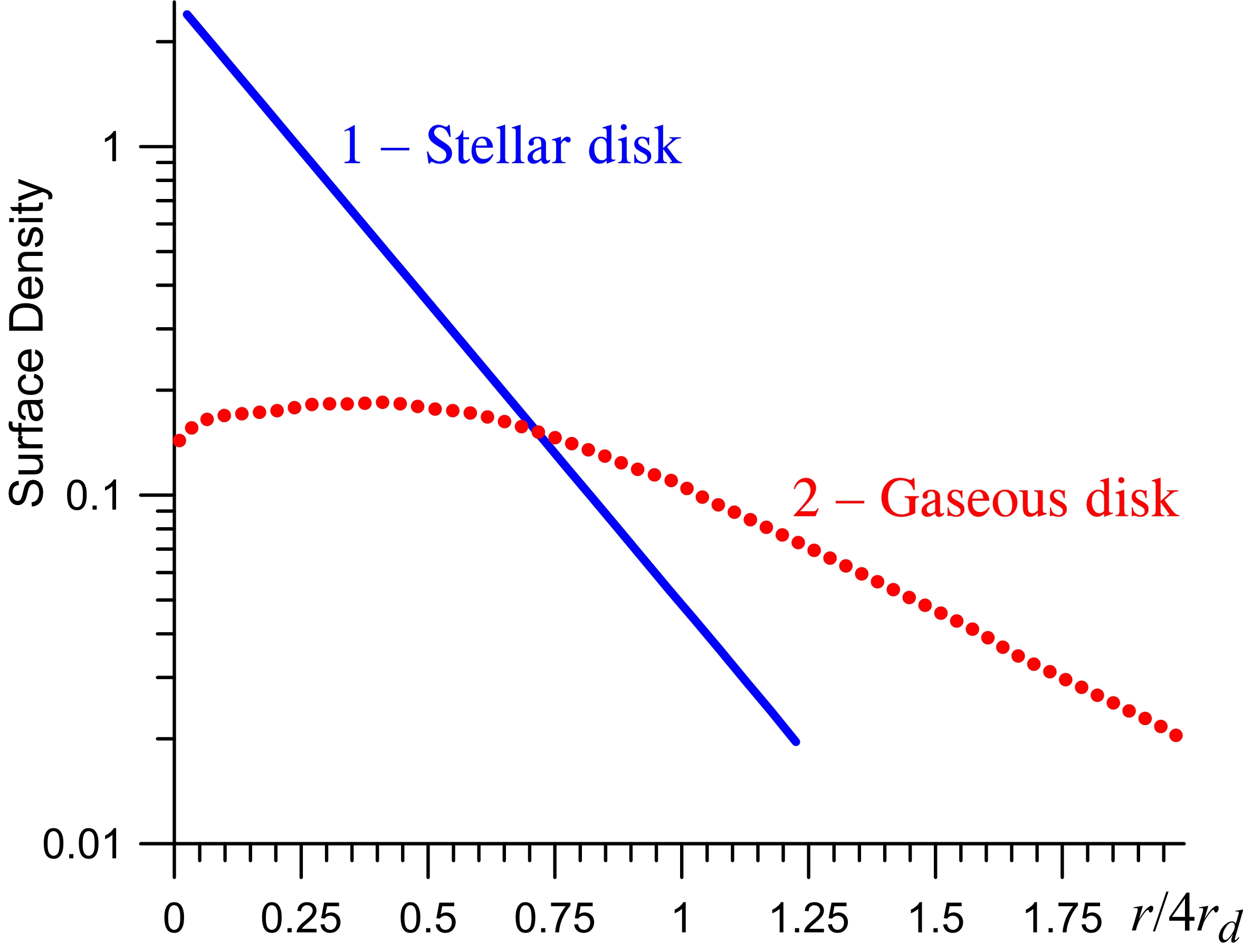}
	\caption{ Initial radial profiles of  stellar surface density (curve 1) and gas surface density (curve 2) for model galaxies, expressed in dimensionless units, where a total mass of stellar component is taken as a unit.
	}
	\label{fig:densityGasStars}
\end{figure}

Radial profile of stellar surface density $\Sigma_*(r)(t=0)$ was approximated by the exponential law with the radial scale-length  $r_d$ (see curve 1 in  Fig.~\ref{fig:densityGasStars}). The shape of initial radial profile of gas surface density $\Sigma_g(r, t=0)$ was taken slowly varying with $r$ with some depression in the central region (curve 2 in Fig.~\ref{fig:densityGasStars}) which is typical for different types of discy  galaxies (See f.e. \citet{xGASS_2020}).  
It was assumed that  gaseous disc extends radially twice as far as a stellar disc. Note that the  results of modelling are not critical to the accepted details of initial radial profile of gas distribution. It is essential, however, that the gaseous disc in the model galaxies is more extended than the stellar one (see Fig.~\ref{fig:densityGasStars}, where $\Sigma_*(r=5r_d) \simeq \Sigma_g(r=8 r_d)$).


In a series of numerical experiments with different initial parameters we trace the evolution of the disc structure, beginning from the initial state of a weak gravitational instability, to a quasi-stationary state reaching after several periods of disc rotation. We use in calculations a dimensionless system of units where a total stellar mass of disc $M_*$ and its initial (``optical'') radius $R=4r_d$ were taken equal to unity. {The circular velocity of a disc at radius R, which is  approximately equal to $GM_{dyn}/R =  (GM_*(1+\mu_g)/R)^{1/2}$, in this system of units is $ (1+\mu_g)^{1/2}$, where $\mu_g = M_g/M_*$.  In a similar way, we find that a period of circular rotation $\tau = 2\pi(1+\mu)^{-1/2}$, and the initial value of the central surface density of stellar disc is $16/2\pi$. }

Note that the density of a model disc, by contrast with masses $M_*$ or $M_g$, is free to  evolve in time in our models. In that  dimensionless system of units, a typical period of revolution of the outer regions  of stellar disc  at the disc radius $R=1$ is $\tau \sim 2.5\div 3$ (depending on the model).  A duration of disc evolution traced in our numerical models ranges from 7~$\tau$ to 15~$\tau$ depending on the  timescale of  perturbations growth. 

The initial parameters of model galaxies are counted below:

\noindent a) Total mass of gas over total stellar mass ratio  $\mu_g = M_g/M_*$.  This value kept constant during the evolution.  

\noindent b) Effective Mach numbers ${\cal M}_* = V_{g}^{\max}/c_r(0)$, and  ${\cal M}_g = V_g^{\max}/c_g$  for stellar and gaseous disc components respectively, $c_g$ being constant along the radial coordinate (see Section 4.1).

\noindent c) A scale-height of stellar disc  $z_0$ at $r\approx 2r_d$ expressed in units of  $r_d$. It is chosen to correspond to the isothermal equilibrium disc after  taking into account a gravitation of gaseous disc and a halo.

\noindent d) Relative mass of  ``rigid'' pseudo-isothermal halo $\mu_h = M_h/M_*$  within the ``optical'' radius $R = 4r_d$,  and the radial halo scale-length $ a> r_d$.

 Note that a live halo would also contribute to the development of disc instability, although it is difficult to expect a significant change in the picture after transition from a rigid to live halo for dwarf galaxies, where  a halo mass usually exceeds a disc mass. Anyway, if spiral arms form in models with a rigid halo, they would develop in the case of a live halo even easier.

\begin{figure}
	\includegraphics[width=\columnwidth]{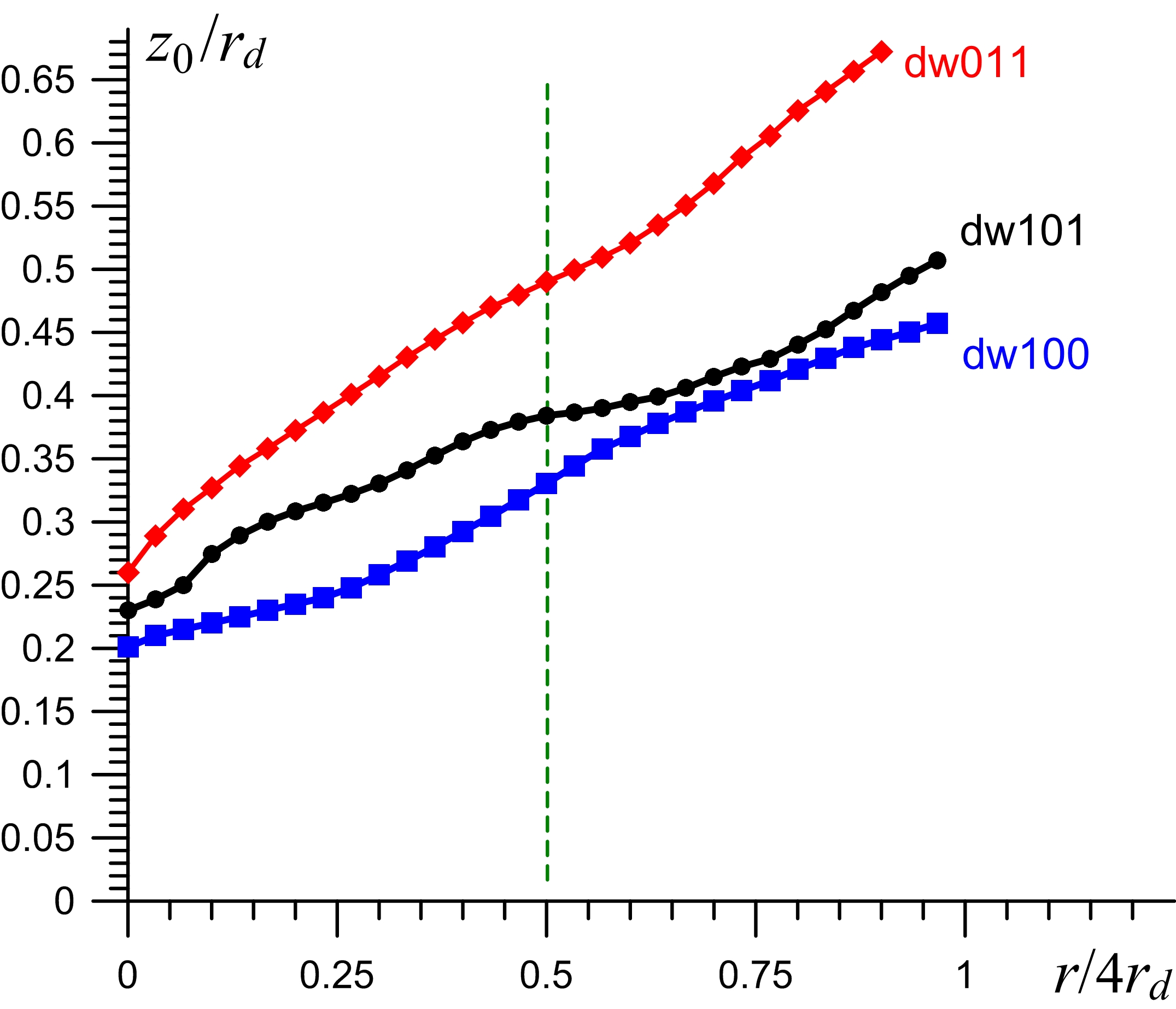}
	\caption{Radial profiles of scale height of stellar dicks for three models. The vertical dashed line marks the doubled radial scale length of discs. 
	}
	\label{fig:VerticalScaleRadius}
\end{figure}

 The initial radial profile of stellar velocity dispersion $c_r(r)$ was assumed to be slowly decreasing along the radius, remaining not lower than $c_g$ everywhere:  $c_r(r) = c_{r0}\exp(-r/r_{dc})$, where $r_{dc}\simeq (6-10)\,r_d$.  
We take a fairly wide range of initial values of  stellar disc thickness and gas content. For the halo component the radial scales  $a\simeq (1.5 - 3)r_d$  and relative masses $\mu_h = 2-3$ were chosen to provide a monotonically growing rotation curve gradually reaching a plateau, which is typical for dwarfs. A choice of effective Mach numbers ${\cal M}_g$ and ${\cal M}_*$  (see above) was dictated by the accepted  constraints for dwarf galaxies: $V_{g}^{\max}\le 100$~km\,sec$^{-1}$ and $c_r > c_g\simeq 10$~km\,sec$^{-1}$.

  {Some remarks should be given concerning the problem of creation of of numerical models of collisionless stellar systems, which is traditionally solved by introducing the cutoff radius $r_c$ in (\ref{eq-gravForcePP}). The collision effect forces a system to evolve toward the relaxation providing the  additional heating of stellar disc due to the kinetic energy of rotation, which is manifested in an increase the velocity dispersion of stars ($c_r, c_\varphi, c_z$)  in the numerical model, accompanied by a slow decrease in the velocity of rotation of a disc. The rate of these changes depends on the cutoff radius  $r_c$ and the number of particles~$N_*$. }

\begin{figure*} 
	\includegraphics[width=0.4\hsize]{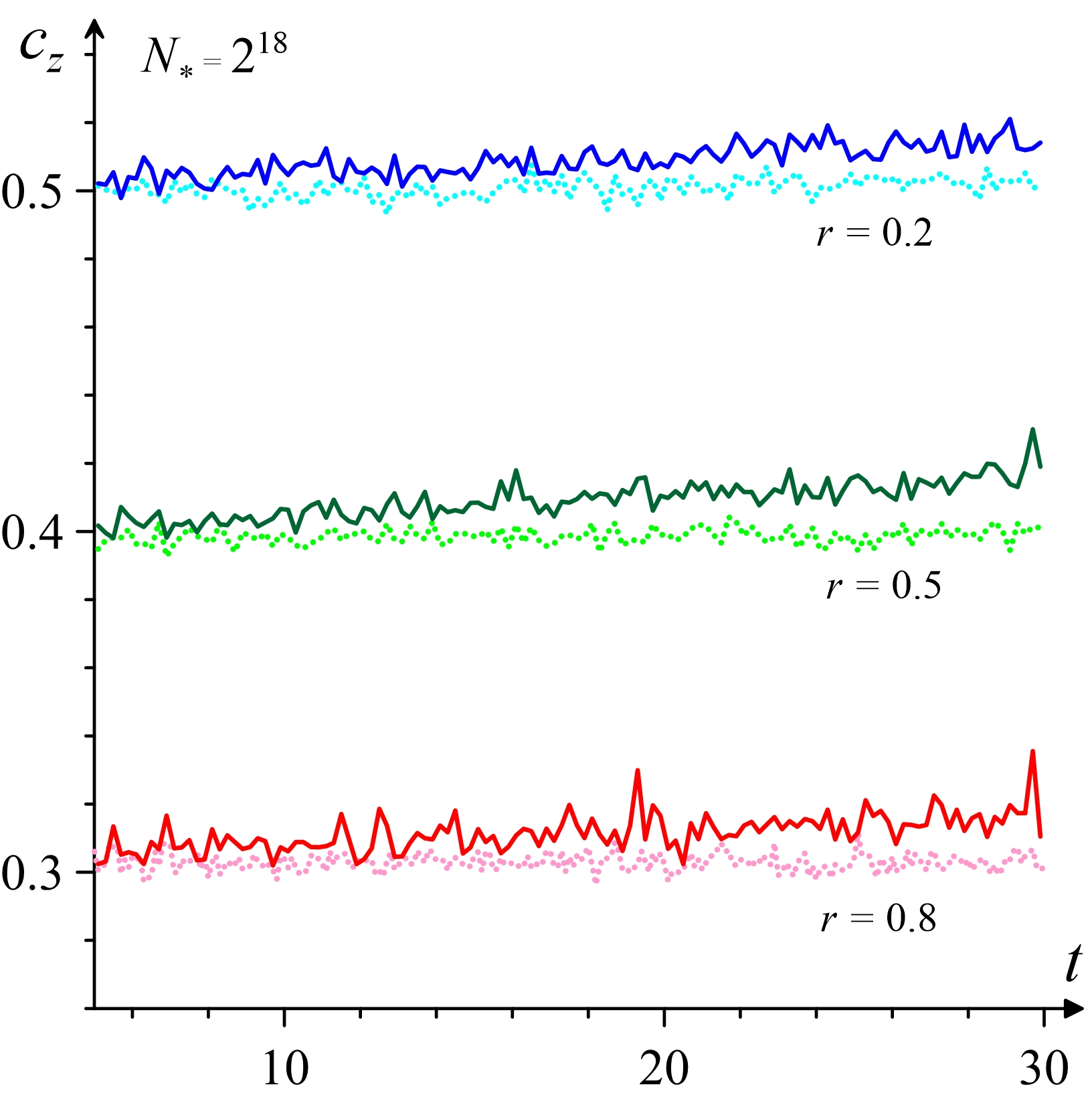} 
	\includegraphics[width=0.4\hsize]{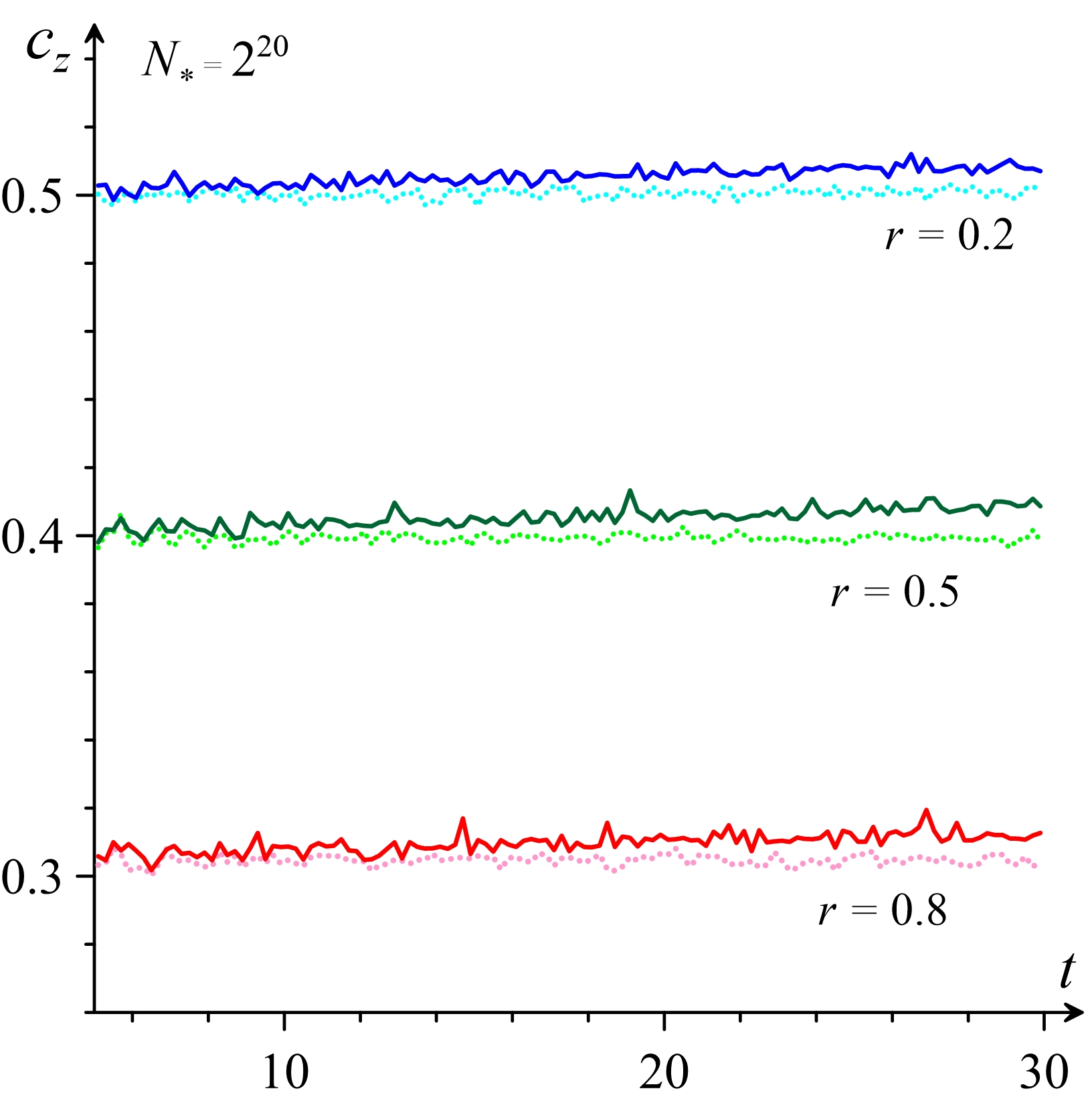}
	\caption{ {Changes with time  of vertical velocity dispersion $c_z$ in the model of stellar disc at three different radial distances: in the central region ($r=0.2$), at two radial scale lengths ($r=0.5$) and at the disc periphery  ($r=0.8$) for two values of the number of points $N_*$ and for two cutoff radii:  $r_c = 0.004$ (dotted lines) and $r_c = 10^{-5}$ (solid lines). }}
	\label{fig:collisionless}
\end{figure*}

{The expected relaxation effects may be demonstrated for a stationary disc in order to exclude strong powerful wave disturbances in a gravitationally unstable system. As an illustration, Figure \ref{fig:collisionless} shows the results of numerical experiments of variation of vertical velocity $c_z$ for a sufficiently hot and stable disc under the different parameters of numerical models (see the legend). The parameter  $r_c = 0.004$  is the basic one for our models, and it ensures the collisionless of the system with the sufficient accuracy. In the case of a small cutoff radii, we have an increase of velocity dispersion and the corresponding   decrease of the velocity of rotation throughout a disc.  As it is expected, with the increase of the number of particles, the heating rate decreases. A detailed discussion on the relaxation problems one may find, for example, in the paper of \citet{Smirnov-Sotnikova2017bar}.} 

All numerical models are self-consistent and begun their evolution from a quasi-equilibrium state. Spiral structure, if it appears, evolves slowly and can persist even longer in the presence of a dynamically cold gas or a strong bar; however, its further fate cannot be predicted with confidence, since it depends on thermodynamic and relaxation processes that are difficult to take into account.

\begin{table}\label{H}
	 	\caption{\label{tab:experiments}
	 	Key  dimensionless parameters of model galaxies: $\mu_g$ is the ratio of gas-to-stellar mass  within the radius $8r_d$; $\mu_h$ is the ratio of halo-to-stellar mass within the conventional ``optical'' radius R = $4r_d$; $a$ is the radial scale length of pseudo-isothermal halo expressed in units of ``optical''  radius; $z_0$ is the scale-height of stellar disc at $r=2r_d$, normalized to an exponential radial scale length $r_d$; ${\cal M}_g$ and ${\cal M}_*$ are the effective  Mach  numbers for gaseous and stellar discs respectively (see the text); column ``Bar'' indicates the presence of a central stellar bar at the end of evolution.
	 	}
\begin{tabular}{lrrrrrrrc}
	& $\mu_g$ & $\mu_h$ & $a$  & $\displaystyle\frac{{z_0}}{r_d}$ &   ${\cal M}_g$ & ${\cal M}_*$ & Bar \\ 
	\hline
	dw001 & 0.01 & 1.94 & 0.50 & 0.43 & 5.47 & 3.27 & -- \\
	dw002 & 0.25 & 1.94 & 0.50 & 0.43 &  6.01 & 3.37 & -- \\
	dw003 & 0.50 & 1.94 & 0.55 & 0.44 & 7.10 & 3.00 & bar \\
	dw006 & 1.00 & 1.94 & 0.4 & 0.45 & 7.50 & 2.90 & -- \\
	dw011 & 1.00 & 1.94 & 0.4 & 0.49 & 7.00  & 3.67 & -- \\
	dw021 & 0.10 & 2.00 & 0.4 & 0.17 & 8.00 & 3.80 & bar \\
	dw029 & 0.40 & 2.00 & 0.4 & 0.26 & 8.95 & 4.42 & -- \\
	dw030 & 0.01 & 3.00 & 0.65 & 0.17 & 8.50 & 4.2 & bar \\
	dw032 & 0.10 & 1.94 & 0.50 & 0.47 & 8.35 & 4.75 & -- \\
	dw033 & 1.00 & 2.45 & 0.65 & 0.36 & 8.5 & 4.0 & -- \\	
	dw041 & 0.10 & 1.94 & 0.50 & 0.26 & 7.40 & 5.35 & bar\\	
	dw042 & 0.20 & 1.94 & 0.50 & 0.29 & 7.71 & 5.48 & bar \\	
	dw043 & 0.40 & 1.94 & 0.50 & 0.33 & 7.95 & 5.53 & -- \\	
	dw044 & 0.60 & 1.94 & 0.50 & 0.39 & 8.14 & 5.66 & -- \\	
	dw045 & 0.80 & 1.94 & 0.50 & 0.41 & 8.83 & 5.77 & -- \\
	dw048 & 0.30 & 1.94 & 0.50 & 0.37 & 7.05 & 5.00 & bar \\
	dw050 & 0.25 & 1.94 & 0.50 & 0.49 & 4.41 & 3.06 & -- \\
	dw053 & 0.60 & 1.94 & 0.50 & 0.51 & 7.76 & 4.92 & -- \\
	dw057 & 0.40 & 1.94 & 0.50 & 0.52 & 5.03 & 2.89 & -- \\
	dw058 & 0.80 & 1.94 & 0.50 & 0.59 & 8.19 & 4.41 & -- \\	
	dw059 & 0.30 & 2.00 & 0.40 & 0.55 & 6.11 & 4.08 & -- \\
	dw061 & 0.45 & 2.00 & 0.40 & 0.58 & 5.20 & 2.82 & -- \\
	dw063 & 0.50 & 2.00 & 0.40 & 0.54 & 5.38 & 2.99 & --  \\	
	dw100 & 0.05 & 3.0 & 0.6 & 0.33 & 8.50 & 4.83 & bar \\
	dw101 & 0.5 & 2.6 & 0.75 & 0.38 & 8.61 & 4.89 & bar \\
	dw102 & 1.0 & 2.3 & 0.75 & 0.39 & 8.71 & 4.95 & oval+ring \\
	dw111 & 0.50 & 2.6 & 0.75 & 0.49 & 8.52 & 4.21 & -- \\
	dw120 & 0.80 & 1.94 & 0.50 & 0.47 & 8.51 & 4.96 & -- \\	
	\hline 
\end{tabular}
\end{table}

Model~discs of galaxies by the character of their evolution can conditionally be divided into three categories: a)~discs which reach the gravitationally stable state after several revolutions remaining strucureless; b) discs, where inhomogeneities of statistical nature slowly developed, however they do not lead to the appearance of a global spiral pattern; c)~discs, where well-defined quasi-stationary spirals developed after 1--2 periods of revolution, which are more pronounced  in gaseous components. Gaseous spiral arms may either look as the Grand Design type, or possess a less ordered structure,  more typical for flocculent galaxies.

The initial parameters of those models where spiral pattern were formed, are given in Table~\ref{tab:experiments}.

\begin{figure*}
	\includegraphics[width=0.999\hsize]{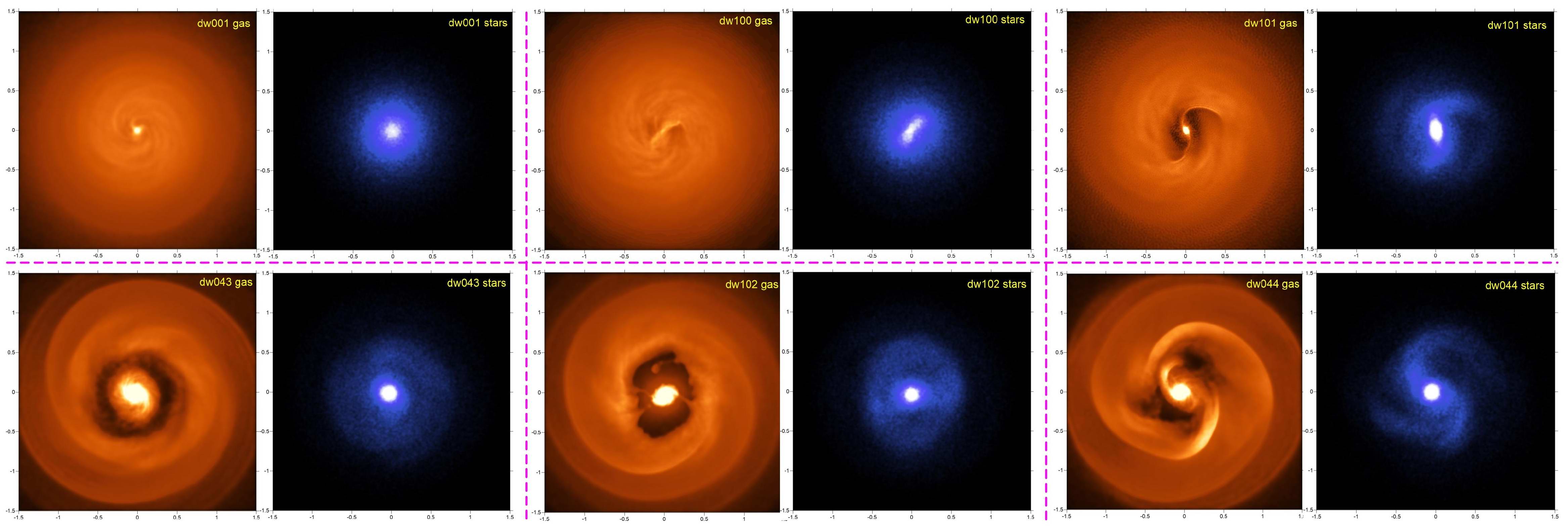}
	\caption{Surface density distribution of gaseous (orange) and stellar (blue) discs for six different models at the end of numerical experiment (see the text). Upper row: model discs where spiral arms in gaseous discs are weak or absent. Note that a  stellar component  may nevertheless reveal a bar or a ring. Bottom row: a well-defined spiral structure formed in the  gas component with shock waves along the spiral arms.}
	\label{fig:spiralStructures}
\end{figure*}

 For illustration, Fig.~\ref{fig:spiralStructures}   demonstrates a general view of the resulting gaseous (orange) and stellar (blue) discs at the end of simulation for six  different initially weakly unstable models. Well defined spirals are seen  in the latter three models (a bottom row).
 The difference of the observed structures is  due to the   differences in gas content and in  the   disc thickness of the model galaxies (see Discussion). 
 
Some model galaxies form a bar, some do not (see Table~\ref{tab:experiments}). As one can expect, the most mature long-lived bars arise in galaxies with less concentrated halos, in which a velocity curve  slowly grows before reaching a plateau. {However, this condition is not very strict for bar formation, and a real situation is more complex even for the bulgeless galaxies we consider.  For example, one of four models (dw021), possessing the most concentrated halos (see Table 3), still reveals a bar, whereas in the model dw111, where the halo is almost twice more extended, a bar is absent. }

\begin{figure*}
	\includegraphics[width=0.999\hsize]{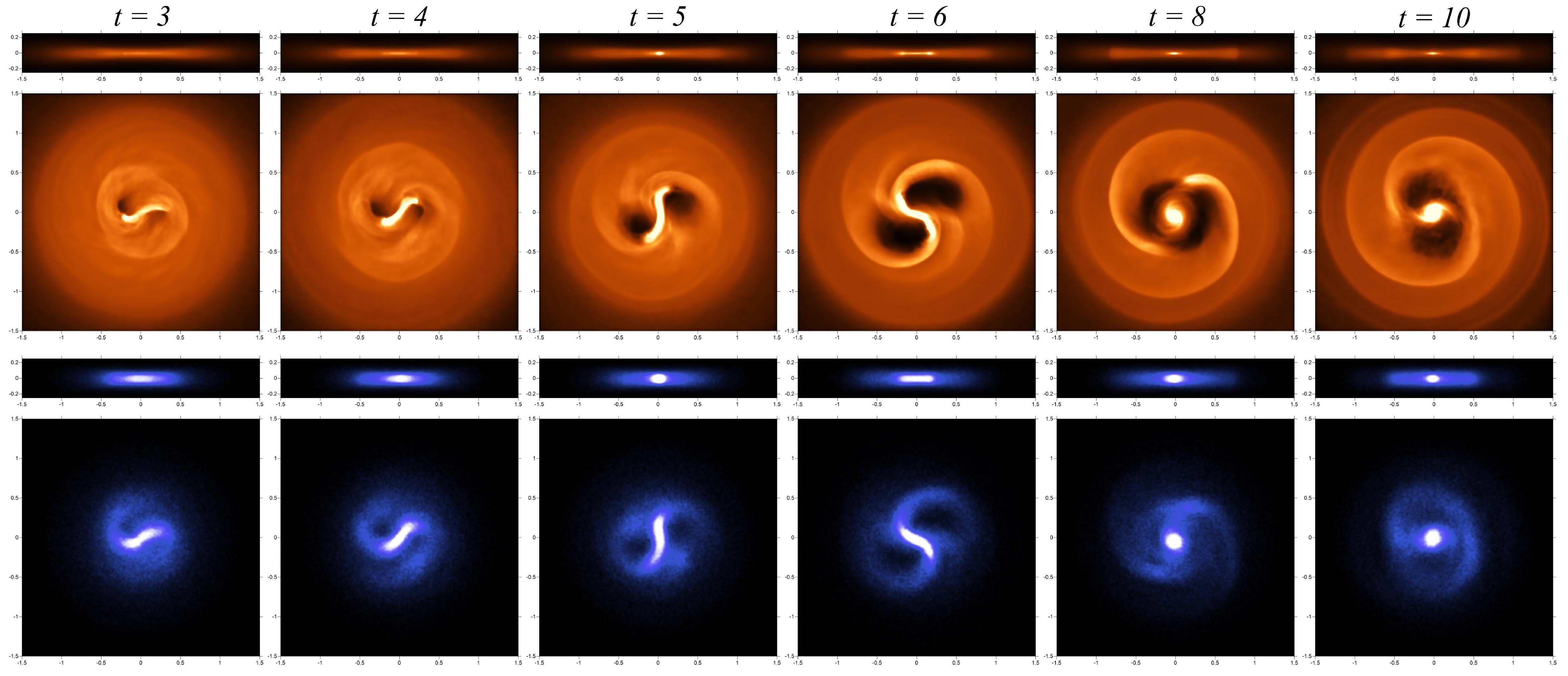}
		\caption{The evolution of gaseous (orange) and stellar (blue) surface densities for model dw102. The time from the beginning of evolution is indicated above the figures. In Fig.  \ref{fig:spiralStructures} the snapshot for this model relates to a later stage of evolution  ($t = 18$). 
}\label{fig:model102bar}
\end{figure*}

{Note that in the models with a large amount of gas, as a rule, the rapid destruction of stellar bar occurs at the  early stages of evolution. A typical example is the dw102 model, where the initial powerful bar initiated a strong radial gas motion leading to the the increase of a central density.This, in turn, led to the transformation of the bar into an oval-like structure with an outer ring (see Fig.~\ref{fig:model102bar}).
A similar situation occured in the dw43 -- dw45 models. In  dw43, one can see a weak oval-like structure in the central zone at the end of evolution, which is a result of destruction of the bar (see Fig.~\ref{fig:spiralStructures}).}  

For a quantitative description  of  spiral arms, a  Fourier analysis of azimuthal distribution of density of particles (gas+stars) was used. Following \citet{Sellwood-Athanassoula-1986},  we calculate the amplitude of the Fourier harmonics of the resulting spiral patterns  as
\begin{equation}\label{4--Eq-Furgarmon}
A(t;p,m)= \frac{1}{ N} \sum\limits_{k=1}^{N} \exp\bigg\{\, i\, \Big[
m\varphi_k(t) + p \ln (r_k(t)) \Big] \bigg\}\,,
\end{equation}
where  
$p$ and~$m$ describe a structure of perturbations along the radial and azimuthal directions respectively \citet{Khrapov-etal-2021Galax}. In particular, ${A}(t;p,2)$ means the amplitude of mode $m=2$ for two-arm structure and a bar. Parameter $p$ describes a pitch angle of spiral arms; $p=0$ for a bar-mode. 

The integral values of  Fourier harmonics at the end of the experiment  may be expressed as 
\begin{equation}\label{4--Eq-intampfur}
\hat{A}(m) = \sqrt{ \sum_{p} |A(p,m)|^2} \,.
\end{equation}
The maximal value of $\hat{A}(t;m)$  reflects a mode number,  that best describes both the shape of a global spiral pattern developed in  a disc and the degree of regularity of the pattern.


\subsection{Toomre' stability parameters}
\label{subsec:NumMode3}
A stability condition for local perturbations of a  disc is usually described by  Toomre' parameter:   $Q_* = c_r/c_T$,  where $c_r$ is a stellar radial velocity dispersion,  and  $c_T = {3.36 G\Sigma_*}/{\varkappa}$ is its local threshold  value  for a thin axisymmetric disc (\citet{Toomre64}).  For a thin disc to be locally  unstable to radial perturbations, it needs to have $Q_* < 1$. Numerical N-body models {taking into account non-radial perturbations and the disc thickness} show that the lower boundary of $Q_*(r)$ for marginally stable collisionless discs of galaxies may vary along the radius,  being higher at the disc periphery (see, for example, \citet{khoperskov-zasov-2003}).
  
 Non-zero disc thickness makes a disc more stable to perturbations.
 {The first to address the stabilizing effect of finite disc thickness was \citet{Vandervoort70} followed by \citet{Morozov-1981stab-stellar, Levi-Morozov1989Ap}. The larger the height scale, the more stable the disc. In the presence of a stellar and gaseous component, the least stable component is decisive for the stability of the disc as a whole.
 }
 
The approximate expressions for modified Toomre' parameter for two-component (stellar and gaseous) discs was given by \citet{Romeo-Falstad-2013} for $Q_g\ge Q_*$:
\begin{equation}\label{eq:defQTsum}
Q_{sum} = \frac{Q_{*}}{1+W Q_{*}/Q_{g}} \,, \quad W = \frac{2c_r c_g}{c_r^2 + c_g^2} \,,
\end{equation}
where $Q_*$ and $Q_g$ are Toomre' parameters taken separately for stellar and gaseous discs respectively. Parameters $Q_*(r)$, $Q_g(r)$ and their generalization $Q_{sum}(r)$ characterize the efficiency of growth of perturbations in gas/stellar discs. The analytic approximation allows to find the most unstable regions of a disc, where  $Q_{sum}$ is minimal, however only numerical models enable to trace a growth of  gravitational instability at non-linear stage, taking into account the radial and ''vertical'' density gradients.
The  analytical expressions for a marginal value of $Q$ for the multiple components of galactic discs 
was considered by several authors (see, for example, \citet{Kim&Ostriker07, Rafikov-2001, Romeo-Falstad-2013} and references therein).

We consider the gravitational instability developed in the disc components as the main mechanism of generation of the observed  spiral-like structures.   The smaller the value of $Q_{sum}$, the more powerful structures one may expect in galactic discs.

\begin{figure} 
	\includegraphics[width=0.8\columnwidth]{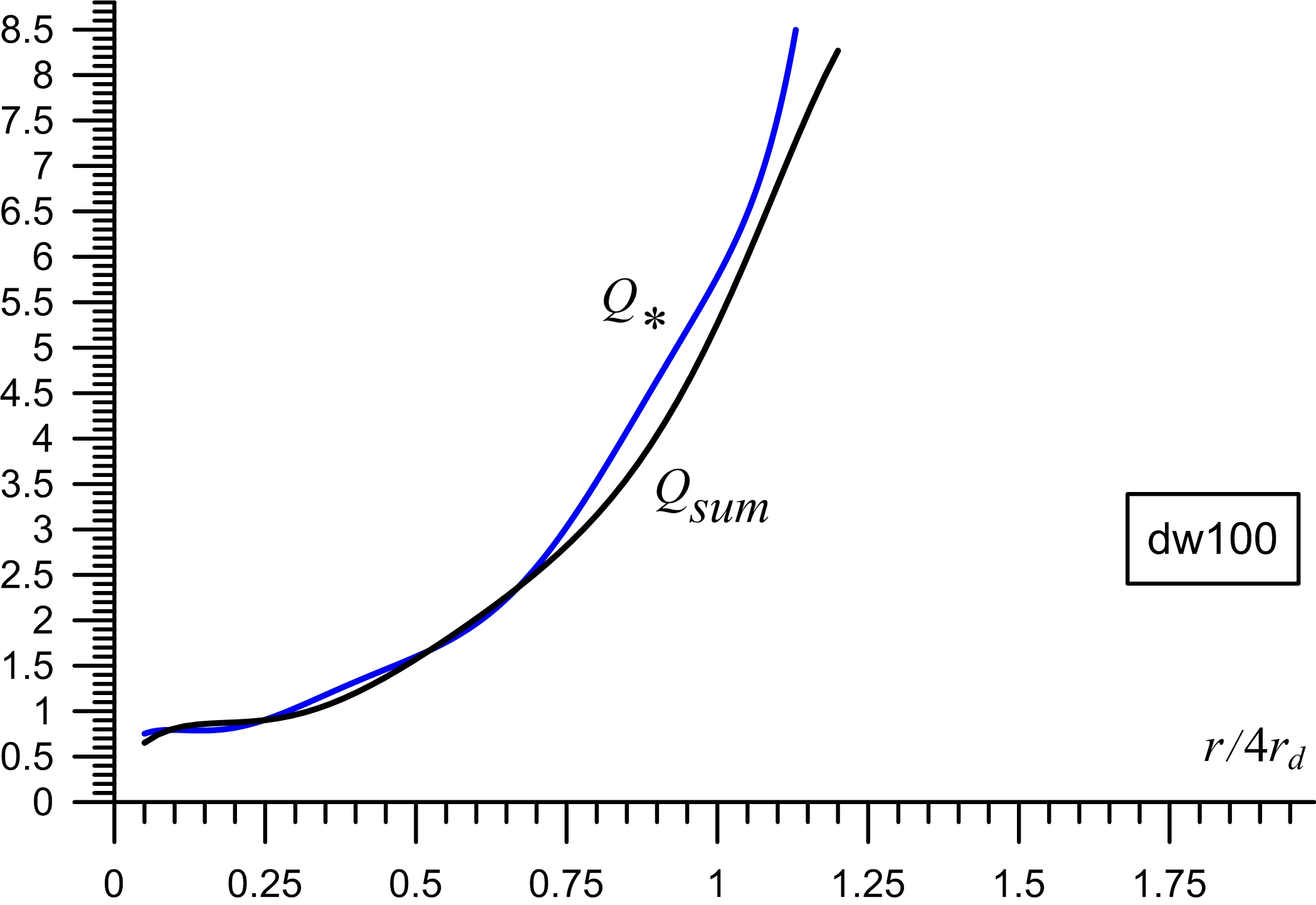} \\
	\includegraphics[width=0.8\columnwidth]{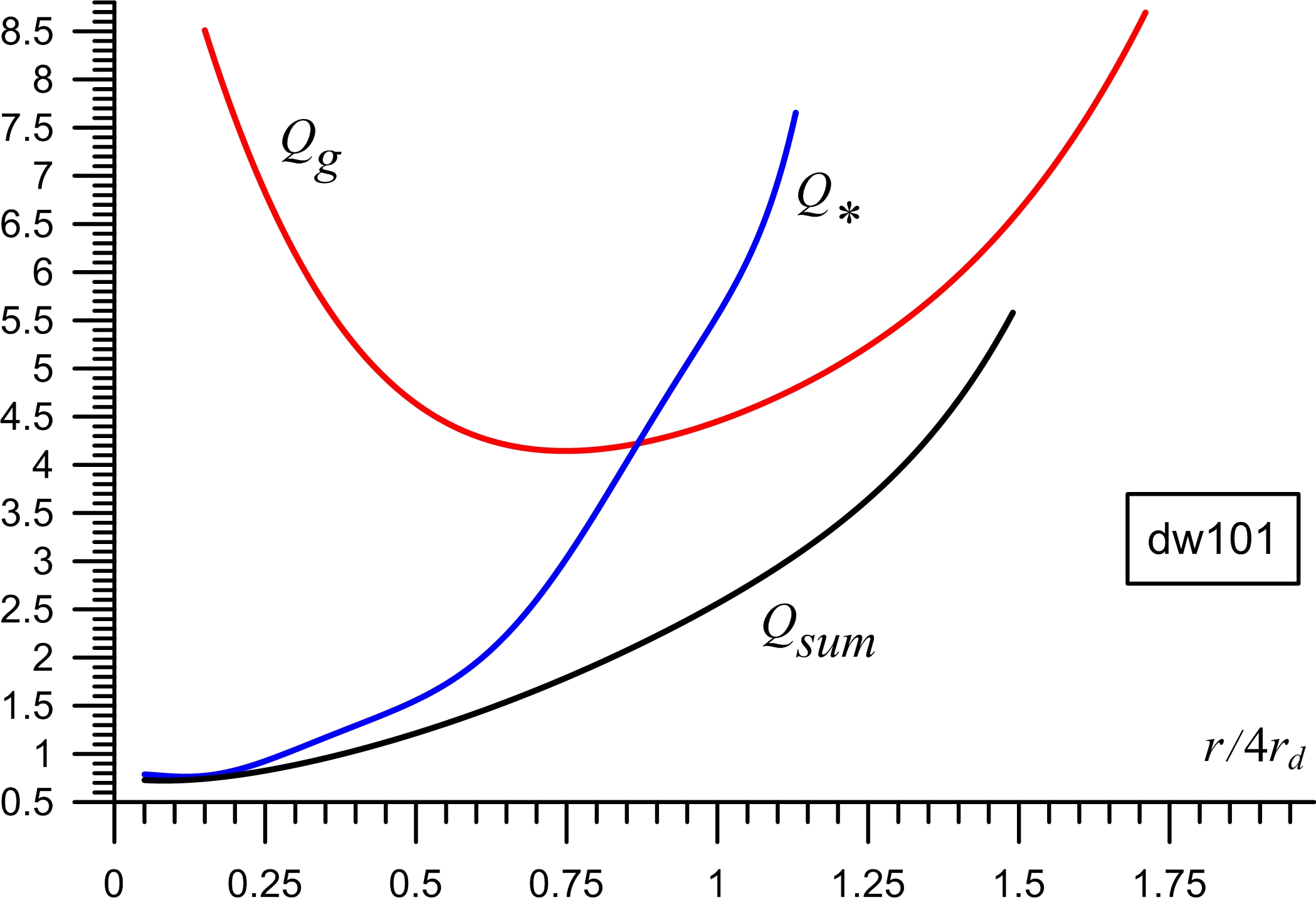} \\
	\includegraphics[width=0.8\columnwidth]{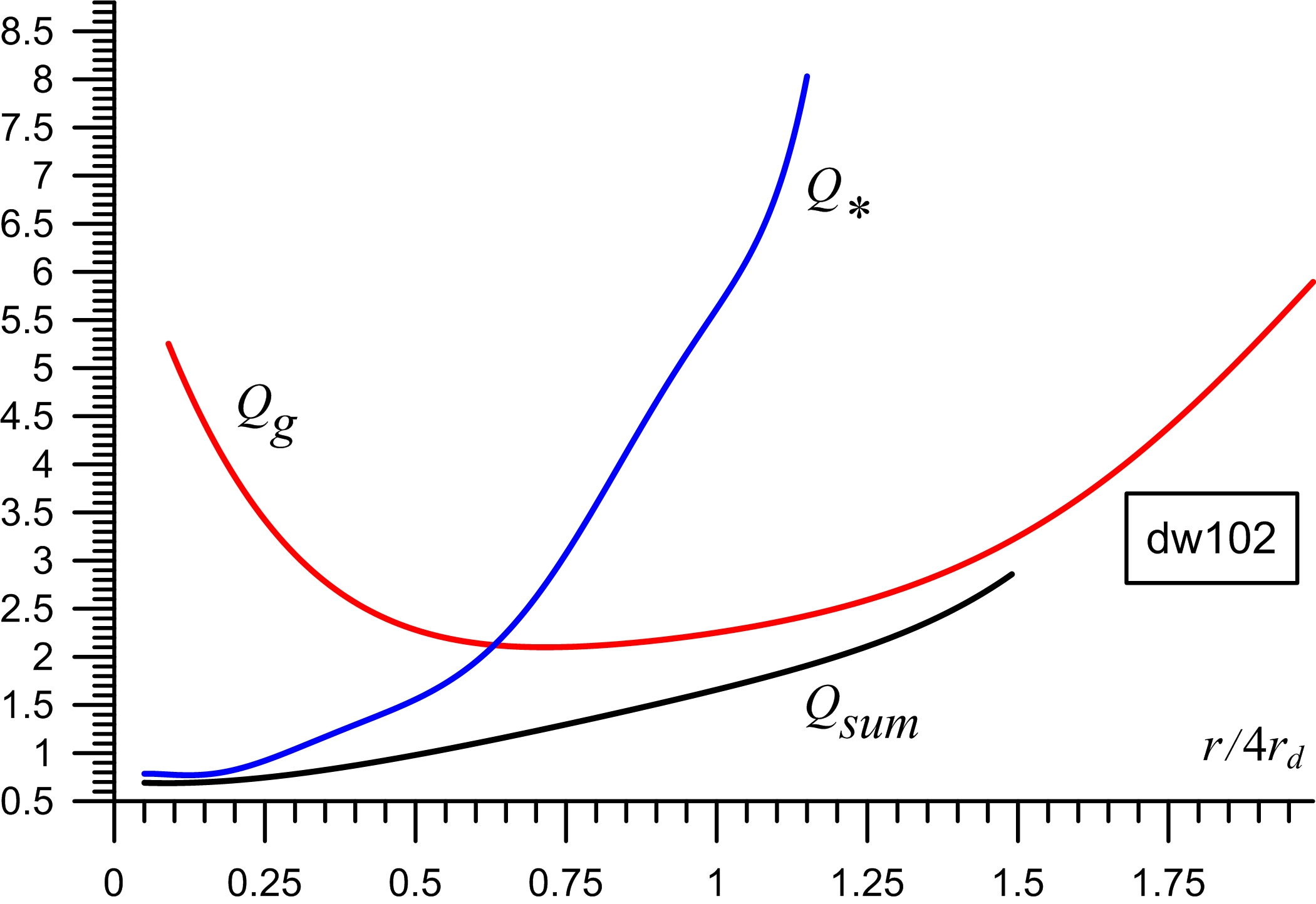}
	\caption{Radial dependences of $Q_*$, $Q_g$ and $Q_{sum}$ at the initial stage of the experiment for three models with  different relative gas mass $\mu_g$: 0.05  (top), 0.5 (middle),  and  1.0 (below). Values of  $Q_{sum}$, indicating the conditions for growth of gravitational instability, are lower where the relative fraction of gas is higher.
}\label{fig:QradialDepend}
\end{figure}

As an illustration, Fig. \ref{fig:QradialDepend} demonstrates the radial profiles  $Q_*(r)$, $Q_g(r)$, $Q_{sum}(r)$ for three models: ``dw100'', ``dw101'' and ``dw102'', which differ by relative mass of gas: $\mu_g \equiv M_g/M_*$\,=\,0.05, 0.5, 1.0 respectively. These models  possess a very similar velocity curves $V_g(r)$, $V_*(r)$ (not shown here), so the difference of radial profiles of Toomre' parameters for them  reflects mostly a different content of cold gas and stellar velocity dispersion of galaxies.

The  radial profiles $Q_{sum}$ shown in Fig. \ref{fig:QradialDepend} are quite typical for the models we consider. They demonstrate that  the best conditions for gravitational instability are located in the central zone of discs $ r\lesssim (1-2) \, r_d $.
A small amount of gas in the model ``dw100'' gives very high values of $ 40 \lesssim Q_g (r) $ (which is beyond the frame of the diagram), with almost no destabilizing effect of gas on the stellar component (Fig. \ref{fig:QradialDepend} \textit{a}).  The increase in   gas mass noticeably reduces $ Q_{sum} $ (Fig.~\ref{fig:QradialDepend} \textit {b, c}), creating the conditions for formation of large-scale spiral pattern despite the fact that for a significant part of the disc the condition of gravitational stability is satisfied.

\section{Discussion and conclusion}
\label{sec:discussion}
A comparison of the samples of spiral (dS) and non-spiral (dSm and dIrr) galaxies led to conclusion that they share similar properties. It  confirms that spiral arms usually occur  in the most massive dwarf galaxies and almost never observed in galaxies whose dynamic mass is less than $10^9~M_\odot$, or if the luminosity $L_B$  is below $3\cdot 10^8\,L_\odot$. There are practically no dS galaxies with rotation velocities below 50--60 km\,sec$^{-1}$. Yet some exceptions still may exist; in our case this is PGC~39837, which luminosity is about $10^8L_\odot$. Note that  its  structure is not typical for dwarfs: it looks like a bar surrounded by a clumpy ring;  such objects deserve more attention.
 
 In terms of isolation, dS-galaxies also do not stand out from other dwarf systems. At the same time, galaxies with an well-ordered structure (Grand Design) do not significantly differ by their properties from those where  spirals have a flocculent type. The presence of bars in dS-galaxies is as common as in non-spiral dwarfs,  which allows to conclude that there is no tight relationship between the formation of a long-lived bar and spiral arms. 
    
The only statistically significant difference we find between dS-galaxies and non-spiral dwarf galaxies with similar characteristics is the lower (on the average) HI content in the former (although a dispersion of gas mass for both types remains high). In the absence of external influences affecting the amount of gas in a galaxy, the most likely explanation for the lower gas content in the galactic disc would be a shorter time-scale of gas consumption, that is, a higher star formation efficiency (star formation per unit of gas mass) $SFE = SFR/M_g$. This assumption agrees with the conclusion of \citet{MaganaSFR20}, where the authors, based on a small sample of late-type galaxies (14 dS and 22 Sm)  inferred that ``dS are forming a larger
number of stars per area than Sm, despite the fact that the latter on average has twice more gas mass than dS''. 

Indeed, a volume density of gas plays a key role in the efficiency of star formation SFE: the lower a density, the lower SFE (see for example \citet{Abramova&Zasov12}, \citet{Bacchini_etal19}). Hence SFE tends to be higher in those galaxies where stellar discs are thinner and denser, since it is a stellar density which  determines a thickness of the equilibrium gas layer in a disc.  Thereby,  one can expect that a more rapid exhaustion of gas takes place in galaxies with thinner stellar discs, other things being equal. On the other hand, thin stellar discs are more unstable to gravitational perturbations, which  promote the formation of a spiral  structure. This may explain the lower (in the mean) gas content in dS-galaxies in comparison with non-spiral dwarfs of similar size, luminosity or dynamic properties (see Section \ref{sec:statistic}).

\begin{figure*}
	\includegraphics[width=0.6\hsize]{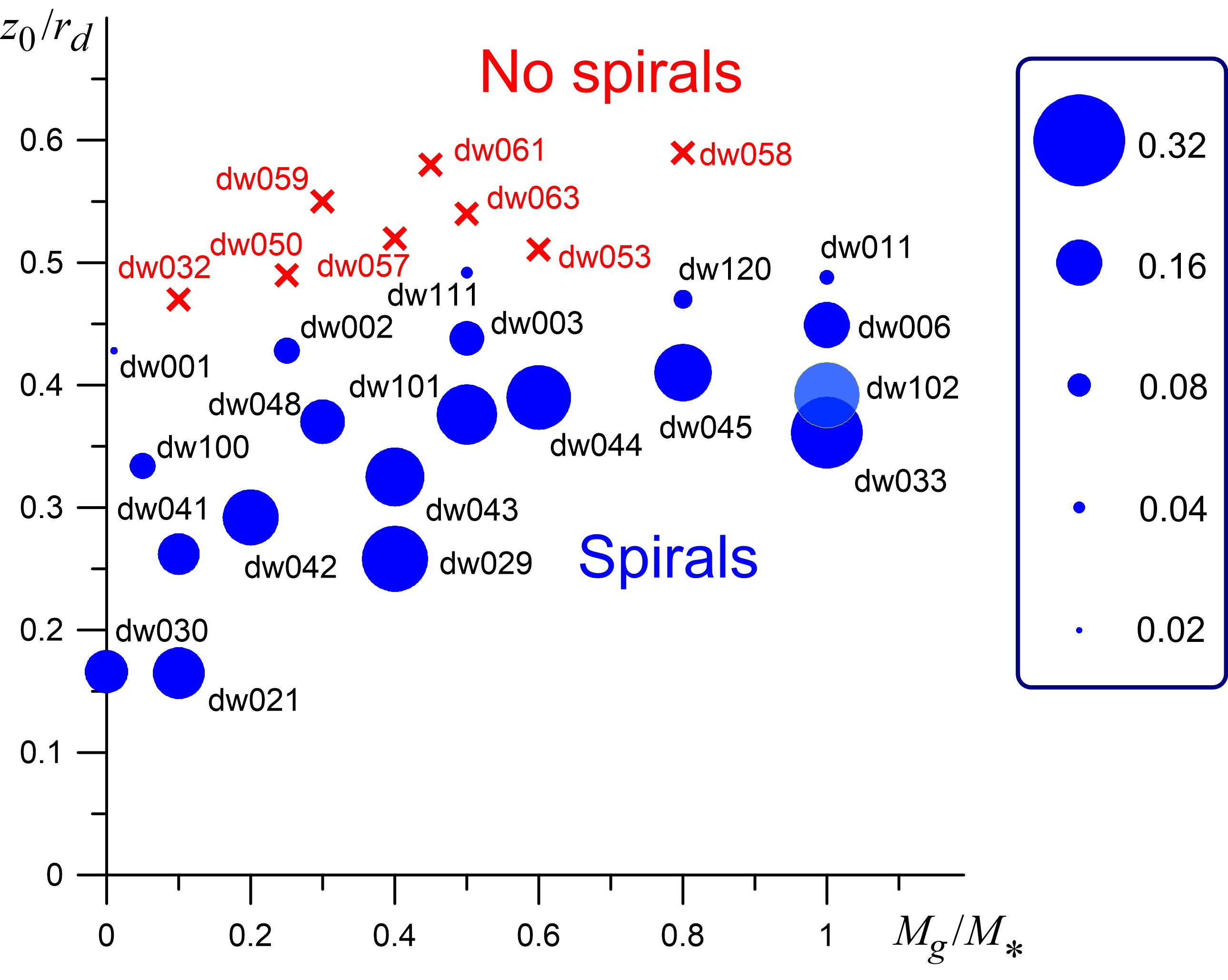}
	\caption{Results of numerical simulations shown in the parameters plane: relative gas mass ($M_g/M_*$) vs  the ratio $ z_0/r_d$, which characterize the relative thickness of a disc. The size of circles characterizes the  maximum Fourier harmonics amplitude of azimuthally averaged density variation of stellar disc, as it follows from (\ref{4--Eq-intampfur}) (see the sidebar). Crosses correspond to models with the absence of inner structure of a disc, which preserved the axisymmetric distribution of parameters of a disc throughout the evolution.}
	\label{fig:FourierHrdMgas}
\end{figure*}

The most favorite conditions for the development of gravitational instability and the formation of extended spiral structure exist in a dynamically cold and relatively thin stellar disc with relatively high velocity of rotation $ V^{max} / c_r $, which corresponds to lower values of the Toomre' parameter. By analyzing the evolution of different models we came to conclusion that in those cases when the maximum gas rotation velocity (a circular velocity) is less than 60~km\,sec$^{-1}$ (or the Mach number $ {\cal M}_g = 6$  if $c_g = 10$\,km\,sec$^{-1}$),  the numerical models fail to form a pronounced global pattern for the range of initial parameters we use.

Diagram in Fig.~\ref{fig:FourierHrdMgas} illustrates a role that the disc thickness and the mass of gas plays in the formation of  spiral structure. The ratio $ z_0/r_d$ at radial distance $r \sim 2r_d$ is compared there with the  ratio of gas-to-stellar mass for different models. Circles relate to those numerical models where the  density waves of different  amplitudes had formed and survived.  A size of circles corresponds to the maximal amplitude of Fourier harmonics  averaged over a disc of model galaxy (see eqs. (\ref{4--Eq-Furgarmon}), (\ref{4--Eq-intampfur})),  obtained by Fourier analysis of azimuthal variation of stellar density. The larger is amplitude, the more contrast and regular is a spiral pattern.  Crosses mark the models where a spiral structure has not formed, and the discs of model galaxies  preserve the axisymmetric distribution of all parameters throughout evolution. 

As it follows from the diagram, spirals may appear even in gas-poor discs, if they are thin enough. The thicker stellar disc, the more gas is required for the spiral structure to form.  All disc models with  $z_0/r_d > 0.45-0.50$ remained structureless during the experiments even when galaxies are rich of dynamically cold gas. 

To conclude,  when a spiral structure is formed, the amplitude of spiral arms depends mainly on two parameters:  first, on the velocity dispersion of stellar population (see Fig.~\ref{fig:FourierHrdMgas}), which determines a relative  thickness of stellar disc, and second, on the relative mass of gas. In addition, a spiral structure does not develop in galaxies of very small size (mass) with a low disc rotation velocity (less than 60~km\,sec$^{-1}$ in our models). The reason for this is evident: the accepted lower limit of stellar velocity dispersion (10~km\,sec$^{-1}$ in our models) causes  low-mass stellar discs to be too fat and stable to develop the inner structure through disc instability. 

It is essential that that in spite of a gas deficiency of dS-galaxies in comparison with dIrr ones, the relative mass of gas in most of dwarf spirals remains high enouth (see Table \ref{tab:spiral}) to allow a spiral structure to develop.

It is evident that the models described above  give us a rather simplified picture of the development  of spiral pattern.  They do not take into account neither star formation nor stellar feedback. It can't be excluded that in some galaxies a regular spiral-like or ring-like features may have a different origin, not connected with the disc self-gravity. It, for example,  may refer to a small slowly rotating  ring galaxy mentioned above --- PGC~39837 (see Table \ref{tab:spiral}).

Although the numerical models under consideration were initially tuned to dwarf galaxies (no bulge, a limitation on the $c_g/V^{max}$ ratios, a   presence of the extended gaseous disc), the diagram in Fig.~\ref{fig:FourierHrdMgas}, which reveals the conditions for development of spiral structure, may have a more general meaning. Indeed, the observations of edge-on galaxies show that the thickness of stellar discs of non-dwarf (spiral) galaxies  lies, with rare exceptions, within the same limits of $0.1\le z_0/r_d\le0.5$  as for our model dS-galaxies  (see f.e. \citet{Mosenkov_et_al15}, \citet{Bizyaev&Mitronova09}, \citet{Sot&Resh&Mos12}, \citet{Zasov_Biz_etal02}). It means that a disc thickness of high luminous  galaxies cover the same  interval, where, judging by the diagram, a spiral structure is expected to be in the presence of a gas component of a disc.

 On the contrary, the axial ratios of Irr galaxies indicate that many of them have fat discs, with fainter galaxies being thicker (\citet{Roychowdhury_et_al13}, \citet{Sanchez-Janssen10}, \citet{Local_TF17}). Based on a joint analysis of photometric and kinematic measurements of a large number of low-luminous galaxies, \citet{Johnson17} came to  conclusion that the intrinsic axial ratios $q_0$,  as well as the ratio $z_0/r_d$  for dIrr galaxies, ranges from 0.1 to 0.8 (note, that  in general case these two parameters, describing a disc thickness,  are not completely equivalent). It agrees with the observations of resolved stellar population of nearby galaxies. The detailed analysis of vertical distribution of stars in six low-mass  edge-on  galaxies observed with the Hubble Space Telescope,  carried out by (\citet{Seth_etal15}) showed that the ratios $z_0/r_d$  for their stellar discs s ranges from 0.16 to 0.56. Moreover,  these galaxies contain  even thicker disc components (see also \citet{Tikhonov06}). Evidently, thick disc components of dwarf galaxies makes them even  more stable.  

Thus, the available data on the thicknesses of stellar discs of low-luminous galaxies and the results of numerical simulations of dynamic evolution of stellar/gaseous discs, described above, confirm that only for galaxies with the most thin discs  the conditions for the formation of spirals are realized. A longevity of spiral pattern  is provided by the  presence of cold gas (some role may also play a bar, formed during the evolution). From this point of view, the observed dS-galaxies are mostly those dwarf systems possessing  a gas component, which   kept a  relative disc thickness below average, that is low enough for a spiral pattern to be formed.

\section*{Acknowledgements}

The authors thank the antonymous referee for fruitful remarks. NZ was supported by Russian Foundation for Basic Research (project 20-02-00080 A). AZ and NZ are grateful the Program of development of M.V. Lomonosov Moscow State University (Leading Scientific School ‘Physics of stars, relativistic objects and galaxies’) for the financial support. 
AK and SK are grateful the Ministry of Science and Higher Education of the Russian Federation (government task No. 0633-2020-0003, all results of numerical simulations of galaxies dynamics in section 4). Numerical calculations were carried by using the equipment of the shared research facilities of HPC computing resources at Volgograd State University.
We acknowledge the usage of the HyperLeda database (\href{http://leda.univ-lyon1.fr}{http://leda.univ-lyon1.fr}).

The Legacy Surveys consist of three individual and complementary projects: the Dark Energy Camera Legacy Survey (DECaLS; NOAO Proposal ID 2014B-0404; PIs: David
Schlegel and Arjun Dey), the Beijing-Arizona Sky Survey
(BASS; NOAO Proposal ID 2015A-0801; PIs: Zhou Xu and
Xiaohui Fan), and the Mayall z-band Legacy Survey (MzLS;
NOAO Proposal ID 2016A-0453; PI: Arjun Dey). DECaLS,
BASS and MzLS together include data obtained, respectively,
at the Blanco telescope, Cerro Tololo Inter-American Observatory, National Optical Astronomy Observatory (NOAO);
the Bok telescope, Steward Observatory, University of Arizona; and the Mayall telescope, Kitt Peak National Observatory, NOAO. The Legacy Surveys project is honored to be
permitted to conduct astronomical research on Iolkam Du’ag
(Kitt Peak), a mountain with particular significance to the
Tohono O’odham Nation.

\bibliographystyle{mnras}

\bibliography{dwarfs.bib} 


\end{document}